\documentclass[12pt,preprint]{aastex}
\usepackage{amsmath}
\usepackage{graphicx}
\usepackage{epstopdf}
\usepackage{subfigure}
\usepackage{morefloats}
\usepackage[normalem]{ulem}

\usepackage[nointegrals]{wasysym}
\usepackage{gensymb}
\usepackage{amssymb}
\usepackage{multirow}
\usepackage{pdflscape}
\usepackage[mathscr]{euscript}

\newcommand{\kms}{\hbox{km~s$^{-1}$}}

\newcommand{\flux}{\hbox{erg~cm$^{-2}$~s$^{-1}$}}

\newcommand{\lumin}{\hbox{erg~s$^{-1}$}}

\newcommand{\be}{\begin{equation}}
\newcommand{\ee}{\end{equation}}
\newcommand{\ba}{\begin{eqnarray}}
\newcommand{\ea}{\end{eqnarray}}

\newcommand{\tele}{Suzaku}
\newcommand{\suz}{Suzaku}
\newcommand{\hea}{\emph{HEAsoft}}

\newcommand{\chan}{Chandra}
\newcommand{\xmm}{XMM-Newton}

\newcommand{\simgt}{\lower 2pt \hbox{$\, \buildrel {\scriptstyle >}\over {\scriptstyle\sim}\,$}}
\newcommand{\simlt}{\lower 2pt \hbox{$\, \buildrel {\scriptstyle <}\over {\scriptstyle\sim}\,$}}
\newcommand{\ls}{\lower 2pt \hbox{$\;\scriptscriptstyle \buildrel<\over\sim\;$}}
\newcommand{\gs}{\lower 2pt \hbox{$\;\scriptscriptstyle \buildrel>\over\sim\;$}}

\newcommand{\pum}{$\pm$}

\newcommand{\rfive}{$r_{500}$}

\providecommand{\e}[1]{\ensuremath{\times 10^{#1}}}

\begin{document}

\def\arcsec{$^{\prime\prime}$}
\def\arcmin{$^{\prime}$}
\def\degr{$^{\circ}$}

\title{Suzaku Measurements of Hot Halo Emission at Outskirts for Two Poor Galaxy Groups: NGC 3402 and NGC 5129 }

\author{Jenna M.\ Nugent\altaffilmark{1}, Xinyu Dai\altaffilmark{1}, Ming Sun\altaffilmark{2}} 

\altaffiltext{1}{Homer L. Dodge Department of Physics and Astronomy,
University of Oklahoma, Norman, OK 73019, USA; jenugent@ou.edu}
\altaffiltext{2}{University of Alabama in Huntsville, Huntsville, AL 35899, USA}

\begin{abstract}
We present \suz\ off-center observations of two poor galaxy groups, NGC~3402 and NGC~5129, with temperatures below 1~keV.  Through spectral decomposition, we measured their surface brightnesses and temperatures out to 530 and 1430 times the critical density of the universe for NGC~3402 and NGC~5129, respectively. These quantities are consistent with extrapolations from existing inner measurements of the two groups. With the refined bolometric X-ray luminosities, both groups prefer $L_X$--$T$ relations without a break in the group regime.
Furthermore, we have determined the electron number densities and hydrostatic masses at these radii. We found that the surface brightness and electron number density profiles require two $\beta$ model components, as well as the indication that a third $\beta$ model may be needed for NGC 3402. Adding the gas mass measured from the X-ray data and stellar mass from group galaxy members, we computed baryon fractions of $f_b$ = 0.0693 $\pm$ 0.0068 and $f_b$ = 0.095 $\pm$ 0.014 for NGC~3402 and NGC~5129, respectively. Combining other poor groups with well-measured X-ray emission to the outskirts, we found an average baryon fraction extrapolated to $r_{500}$ of $\overline{f_{b,500}}$ = 0.0912 $\pm$ 0.0050 for X-ray-bright groups with temperatures between 0.8 and 1.3~keV, extending existing constraints to lower-mass systems and indicating that significant baryon losses exist below approximately $r_{500}$.

\end{abstract}


\section{Introduction} \label{sec:intro}

Galaxy clusters and groups are virialized overdensity regions in the universe. Based on numerical or semianalytical simulations (e.g., \citealt{bn98}), the overdensity of clusters and groups in the virial radius, $r_{vir}$, is approximately 100 times the critical density of the universe for the prevailing concordance cosmology ($r_{100}$).
However, observations are more easily able to probe the central regions within $\sim r_{2500}$, which limits us from understanding the overall properties of these objects, such as their virial masses, temperatures, and gas and stellar contents.
Therefore, measuring cluster and group properties at their outskirts close to the virial radius becomes a major endeavor.   
For galaxy clusters, successful measurements of the X-ray emission near $r_{200}$ have been made with \suz\ for many individual clusters \citep[e.g.,][]{fuj08,bau09, geo09, rei09, hos10, kawa10, aka11, sim11, sat12,walk12,ichi13} and by using stacking analysis \citep{dai07,ryk08,shen08,dai10,eck12}.  Yet for galaxy groups, it is more difficult to study the X-ray emission at large radii because of the relatively weaker emission.  The situation is especially severe for poor groups with temperatures below $k_BT \lesssim 1$~keV, where only measurements from stacking analysis and very few individual systems exist for these groups \citep{dai07,sun09,and15}. 

Galaxy groups are important to study the properties of virialized structures, especially to test the deviations from self-similar model predictions, such as the $L_X$--$T$ relation. Scaling relations are extremely useful for better understanding the physics of various types of objects and can be used to perform simulations, e.g.,~\citet{tro18} and \citet{krav06}. Since there are very few measurements out to large radii for poor groups, the nature of these relations at lower mass ranges is not well known. More accurate measurements in the group regime will extend the mass range for these tests.

Groups of galaxies are also important to better quantify the missing baryon problem in the low-redshift universe (\citealt{breg18} and references therein),  
in which the observed amount of baryons is less than that determined based on the cosmic microwave background (CMB) observed from the early universe.
We know this by comparing the fraction of baryonic to total matter that has been obtained for both high-redshift CMB studies and nearby, low-redshift surveys of galaxies, galaxy groups, and clusters. According to the 3 yr WMAP data, which we assume in this work, the baryon fraction, or the ratio of baryonic to total gravitational matter, is $f_b=\Omega_b/\Omega_m = 0.175$ for $\Omega_m = 0.26$ and $\Omega_\Lambda=0.74$. While observations of nearby galaxies yielded only about 10\% of the expected baryon content \citep[e.g.,][]{pers92,bris94,fuku98}, observations of rich galaxy clusters with $k_BT>$ 5~keV retain the cosmological value after adjusting for stellar mass \citep{vik06}.

Illustrated by Figure~\ref{fig:bfrac} in the Discussion, we can see that the observed baryon fraction of nearby systems increases as a function of gravitational potential well and follows a broken power-law model \citep{dai10,dai12}. The data for all but the most massive objects fall below the cosmological fraction measured at high redshift. 
The group regime is arguably the transition region, where baryon loss becomes significant. However, we lack sufficient data to well define the mass threshold of the baryon loss owing to difficulties in accurately measuring their properties, especially to the outskirts.
These missing baryons are theorized to be in a warm-hot intergalactic medium, which permeates the large-scale structure filaments of the universe and hot gas halos of galaxy clusters and groups. Recent work has strongly supported this hypothesis \citep[e.g.,][]{nico18}.

Though this general picture is likely correct, some key questions still remain ambiguous, such as how virialized regions of various galactic systems lose their baryons or if the missing baryons of galaxies and galaxy groups were suppressed from falling into their hot gaseous halos altogether.
Answering these questions will guide the development of numerical simulations with nongravitational processes such as feedback and preheating \citep[e.g.,][]{bens10}.

In this work, we observed the diffuse, extended emission from two poor galaxy groups in the soft X-ray band with \suz, which is best for such observations due to its low, stable background resulting from its low Earth orbit. The two groups extensively studied in this paper and many of their properties are well documented in the literature. For instance, NGC 3402 Group, also called SS2b153, NGC 3411 Group, and USGC S152, appears to be perfectly round, containing ``no evidence of irregularity" \citep{mah05}, and is believed to have between four and five member galaxies \citep{mah05,guz09}. This nearby ($z=0.0153$) group has a global temperature $k_BT=0.88 \pm 0.04$ keV \citep{sun09}. All global temperatures mentioned in this work have been adjusted for the significant change in AtomDB, as discussed in Section~\ref{sec:temp}. NGC 3402 is the central giant elliptical galaxy of this group, classified as a cD galaxy. Accordingly, the group has been labeled a fossil group, which is believed to be the remnant of a series of galaxy mergers resulting in a dominant, bright elliptical galaxy surrounded by a few, much less luminous galaxies \citep{jones03}. 

Although NGC 5129 Group has nearly the same global temperature as NGC 3402 Group, $k_BT=0.90 \pm 0.04$ keV \citep{sun09}, it is a less nearby ($z=0.0230$), loose group with approximately 19 member galaxies \citep{mah04}. The term ``loose" means that the galaxies are separated by greater than several galactic radii on the sky \citep{hel00}. Though different in some ways, both groups lie in the temperature range that so far has a dearth of successful measurements. This is especially true for their outskirts, hence the need for our \suz~observations. 

Throughout this paper, we adopt the 3 yr WMAP cosmology and a flat universe: $H_0$ = 73 km s$^{-1}$ Mpc$^{-1}$, $\Omega_m$ = 0.26, and $\Omega_\Lambda$ = 0.74. Beyond this Introduction, Section~\ref{sec:obs} provides details on the observations, as well as the extensive data reduction we performed on these groups. Next, the process by which we determined the surface brightnesses (SBs) through spectral analysis is detailed in Section~\ref{sec:surf}. Also, numerous radial profiles are plotted and the process of obtaining our contribution to them is explained in Section~\ref{sec:rad}. Moreover, several mass quantities later used to determine various mass fractions are established in Section~\ref{sec:mass}. Finally, in Section~\ref{sec:disc} we discuss our findings and summarize their larger implications.

\section {Observations and Data Reduction} \label{sec:obs}
Observations were obtained of NGC~3402 Group (hereafter NGC 3402), centered at 22.$^\prime$1 ($r_{475}$) away from the group X-ray center with a position angle (PA) of 108$^{\circ}$ in the X-ray band using \suz\ on 2010 December 27 for 49 ks.
Also, we observed NGC~5129 Group (hereafter NGC 5129), using two off-center pointings with separations from the group X-ray center of 16.$^\prime$2 and 15.$^\prime$3 (farther at $r_{420}$). This was performed on 2010 December 18 with PAs of 78$^{\circ}$ and 161$^{\circ}$ and raw exposure times of 55 ks and 38 ks, respectively. These off-center observations are also referred to as target or outskirts observations in this work. Additionally, to better model the background, we performed one background pointing for each galaxy group at $2.04r_{200}$ and $2.14r_{200}$ for NGC 3402 and NGC 5129, respectively.  
The two background observations were carried out within 10 days of the corresponding target observations to ensure that no significant time variability had occurred in the X-ray background between them.

\begin{figure}[h]
	\includegraphics[width=.5\textwidth]{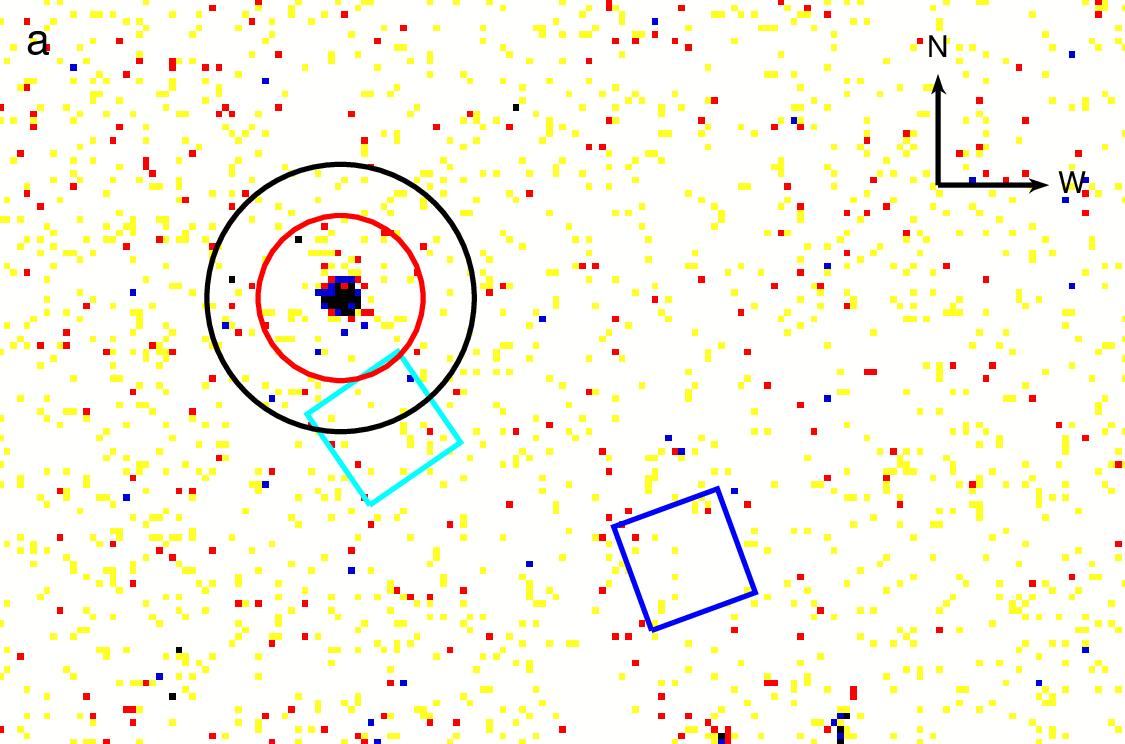} 
	\includegraphics[width=.5\textwidth]{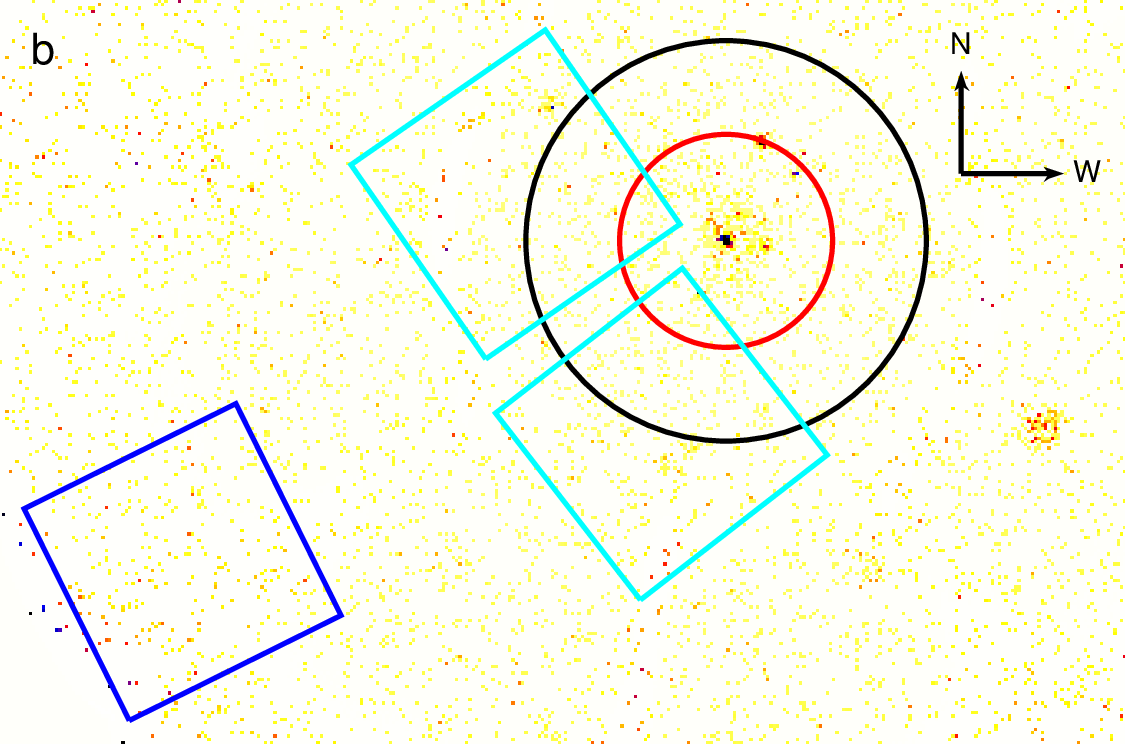}
	\caption{ROSAT images for (a) NGC 3402 and (b) NGC 5129 with overlaid extent of the \chan~spectral analysis from \citet{sun09} (red circles), extent of $r_{500}$ according to the electron number density profiles determined in this work (black circles), the \suz\ FOV for observations of the two groups (cyan squares), and their corresponding \suz\ background observations (blue squares). The cyan and blue squares are 17.$^\prime$8 on each side.}
	\label{fig:rosat}
\end{figure}

All five observations were taken using the three remaining X-ray Imaging Spectrometers (XISs) on board \suz: two front-illuminated (FI) CCDs (XIS0 and XIS3) and one back-illuminated (BI) CCD (XIS1).  
Details of these observations are listed in Table~\ref{tab:obspar}. Also, ROSAT images of each group are depicted in Figure~\ref{fig:rosat}, where the radial extent of the \chan\ analysis from \citet{sun09}, the extent of $r_{500}$ based on the electron number density profiles discussed later in this work, and the \suz\ field of view (FOV) for the group and background observations are shown. In panel (b) of Figure~\ref{fig:rosat}, the northern observation is what we have designated NGC 5129 1st, whereas the southern observation is NGC 5129 2nd. From this, we can see that the center of each group pointing lies beyond $r_{500}$ and a significant area of NGC 5129 is analyzed owing to its two spatially separate pointings. 

The data were reduced using the software package \hea\ version 6.13. 
First, we reprocessed the data using the FTOOL \verb+aepipeline+, which also performs default screening, along with the XIS calibration database (20120210). 
All data were reduced according to the Suzaku Data Reduction Guide.\footnote[3]{http://heasarc.gsfc.nasa.gov/docs/suzaku/analysis/abc/abc.html} Additionally, we excluded times when the revised cutoff rigidity value (COR2) was less than 6 GV to improve the signal-to-noise ratio (S/N) by reducing instances of background flaring. These are times in which the satellite passes through regions where the geomagnetic field is weak \citep{tawa08}.

\begin{figure}[h]
	\includegraphics[width=.45\textwidth]{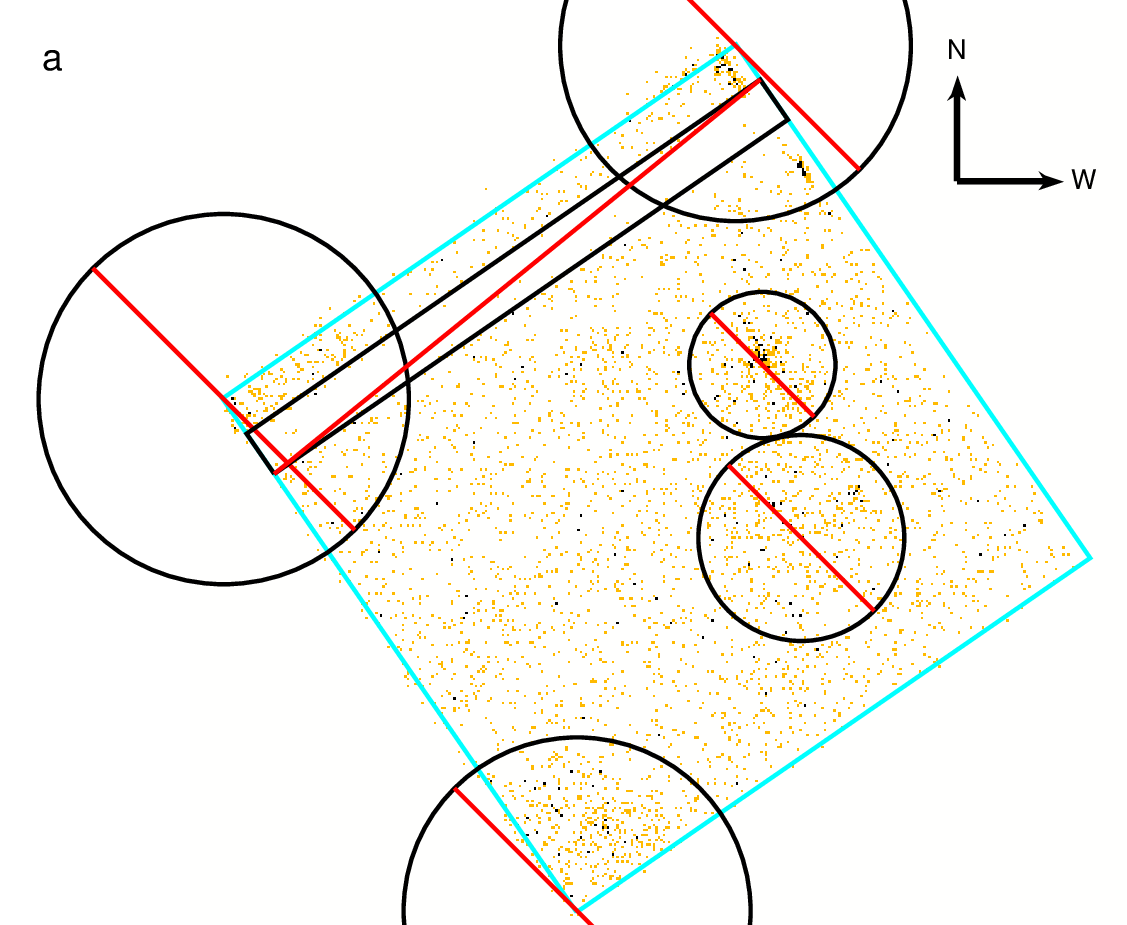}
	\includegraphics[width=.55\textwidth]{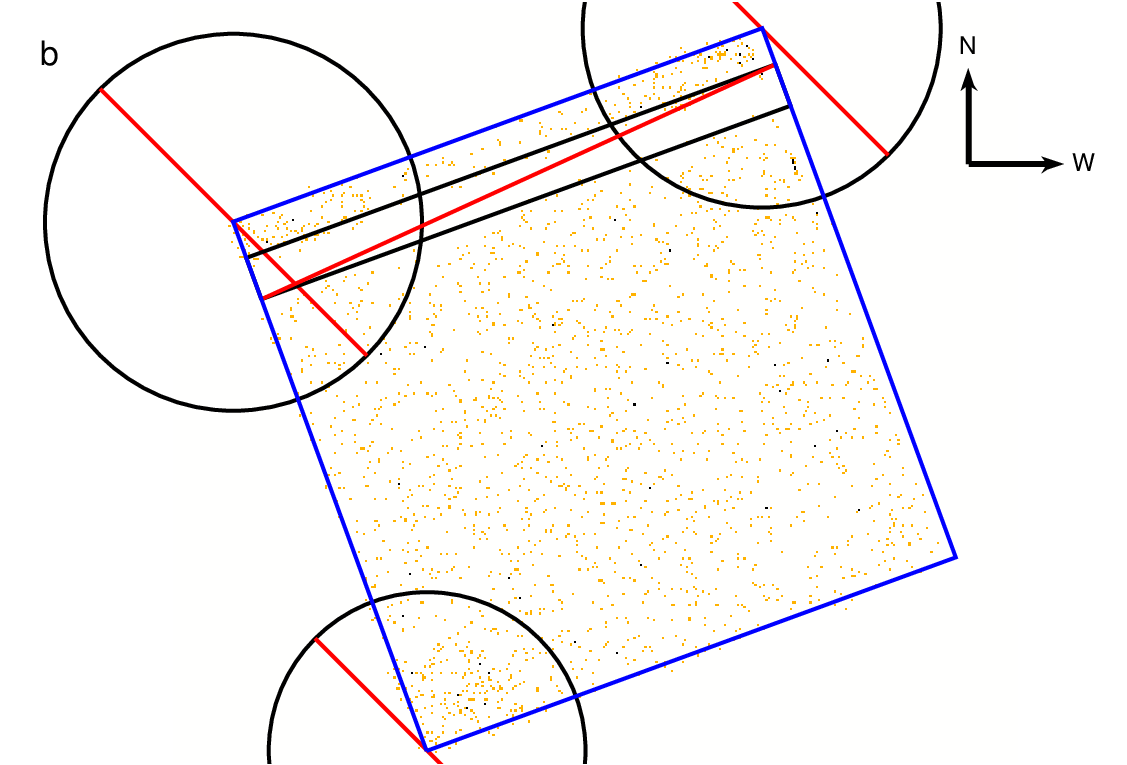}
	\includegraphics[width=.5\textwidth]{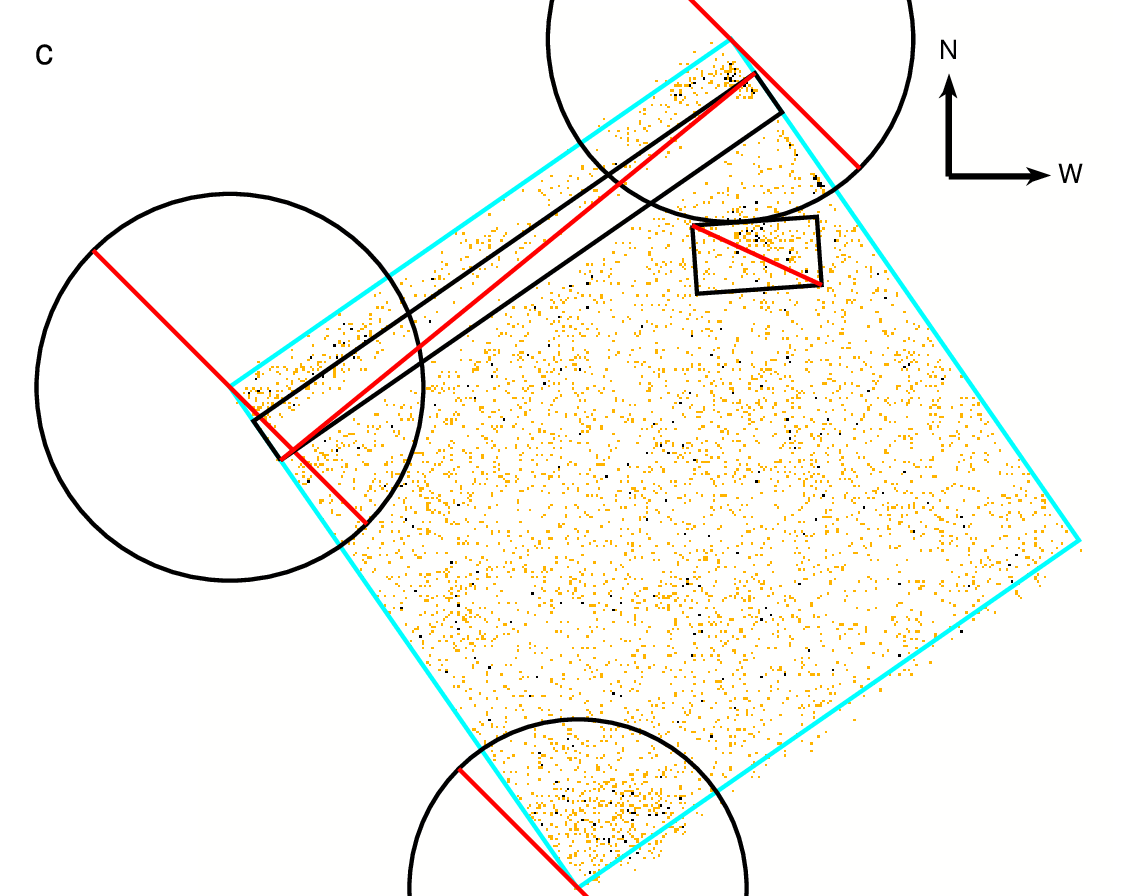}
	\includegraphics[width=.5\textwidth]{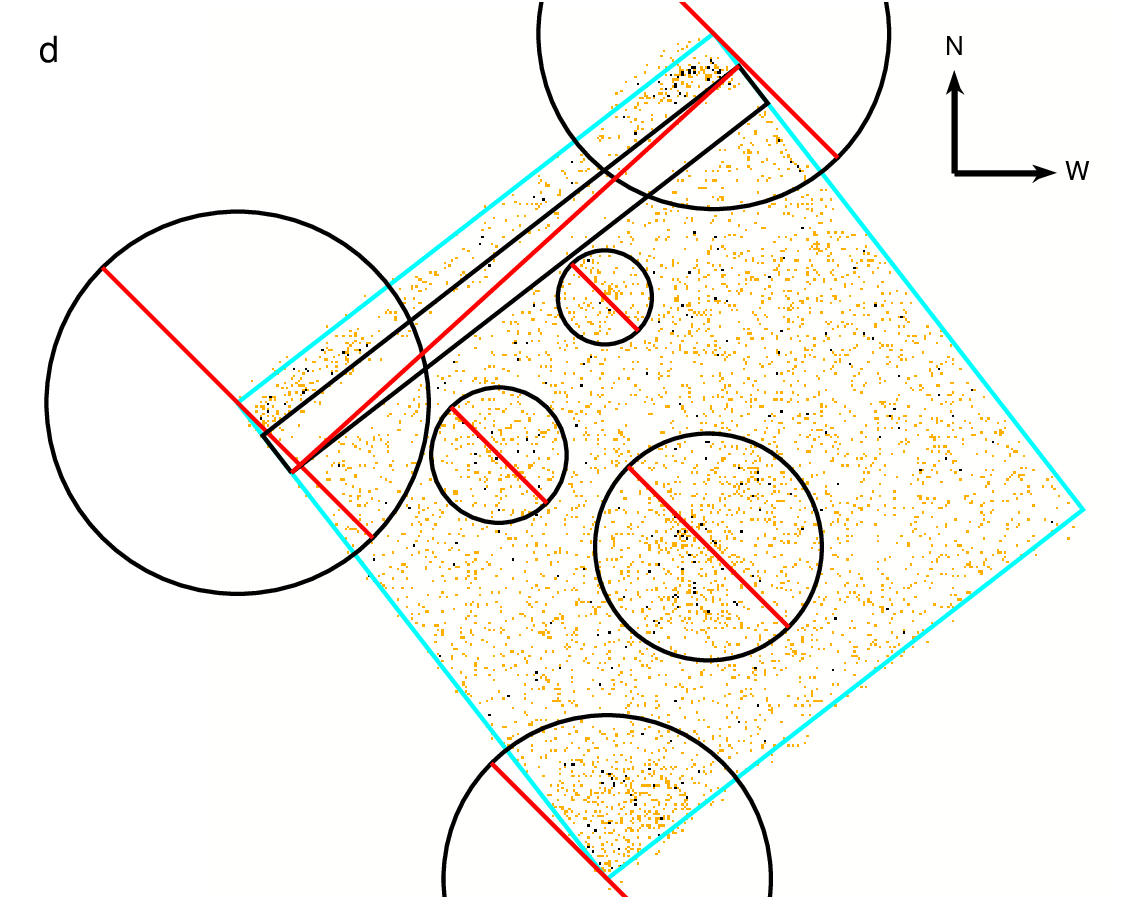}
	\centerline{\includegraphics[width=.5\textwidth]{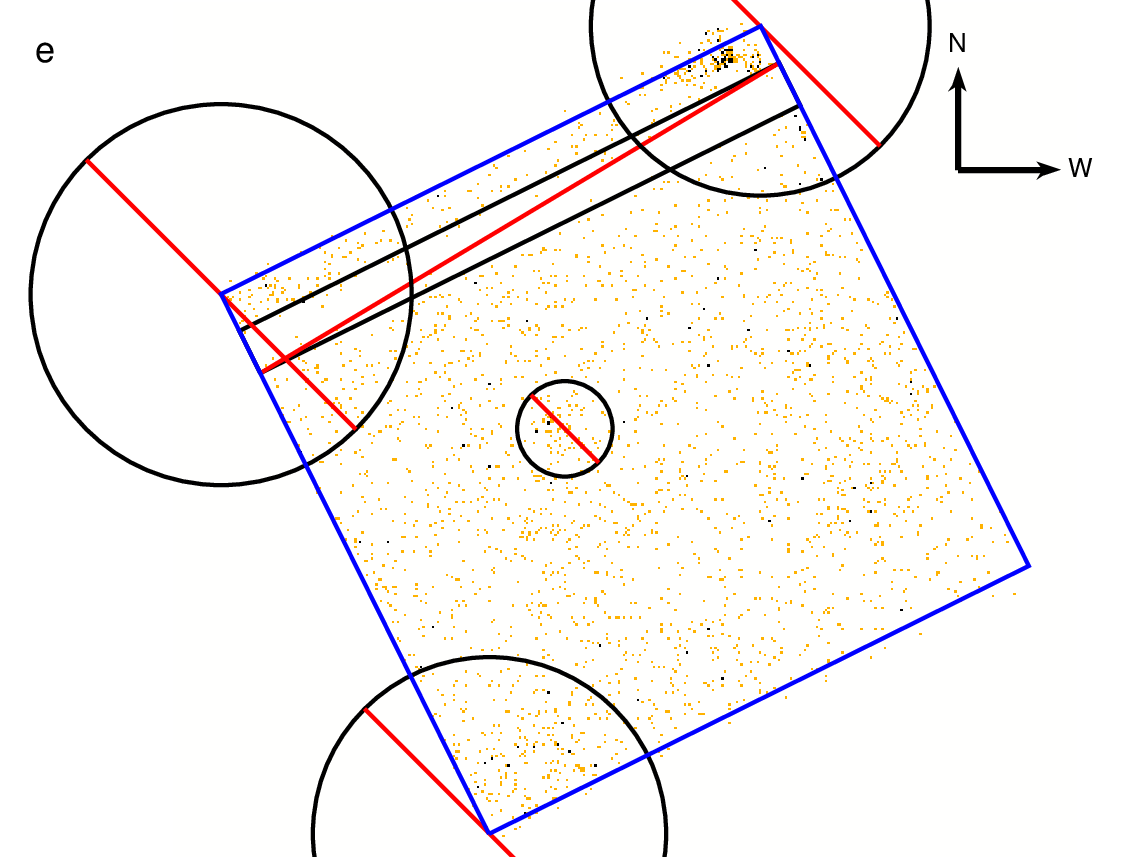}}
	\caption{\suz~XIS0 3x3 and 5x5 combined formatted images for (a) NGC 3402 outskirts, (b) NGC 3402 background, (c) NGC 5129 1st outskirts, (d) NGC 5129 2nd outskirts, and (e) NGC 5129 background pointings with inclusion and exclusion regions and the COR2 $>$ 6 GV condition applied.}
	\label{fig:obs}
\end{figure}

Then, we removed resolved foreground and background X-ray sources, as well as the $^{55}$Fe calibration sources located at two corners of each detector (Figure~\ref{fig:obs}). 
The locations of the calibration sources were known, and the remaining sources were excised by visual inspection. 
Furthermore, most likely due to a micrometeorite impact, a strip of the XIS0 detector (located at DETX = 70--150) was deemed unusable by the XIS team.\footnote[4]{http://www.astro.isas.ac.jp/suzaku/doc/suzakumemo/suzakumemo-2010-01.pdf} Following their notes for reducing XIS0 data after this anomaly,\footnote[5]{http://www.astro.isas.jaxa.jp/suzaku/analysis/xis/xis0\_area\_discriminaion} we used a C-shell script to generate a region to remove all events in the affected area and formed a region to remove possible spurious sources near this strip. 
This was applied to the XIS0 CCDs for all observations. Figure~\ref{fig:obs} illustrates these sources and their regions for the XIS0 3x3 and 5x5 combined observations. 

Next, we examined these observations' light curves using \verb+Xselect+ for instances of background flaring in the 0.5--8~keV band after the above screening processes. 
The only light curves that seemed to indicate any flaring were from the BI CCD (XIS1). Since our analysis is focused solely on the soft X-ray band, we filtered those spectra further by the energy band used later on in the spectral analysis: 0.5--5 keV \citep{ichi13}. The resultant curves show no flaring in that energy range. Thus, the light curves in the energy bands of interest are not contaminated by background flaring. Therefore, no significant background flares were found in any observations.

\section{Surface Brightness} \label{sec:surf}
Two methods were employed to measure the mean SB for each target. First is the direct subtraction method, since there are background observations at greater than $2.04r_{200}$ performed within 10 days of each target observation enabling the non-X-ray background (NXB) to be measured well. In this method, the SB is computed for both the target and their corresponding background observations, and the net value is the difference between the two.
The second method involves modeling the spectra of both the target and background observations, and the SB is determined from the best-fit model parameters for the group emission.
\subsection{Direct Subtraction Method} \label{sec:dirsub}

Using version 2.4b of \verb+Xselect+, we read in both the 3x3 and 5x5 event files with the COR2 $>$ 6 GV screening for each CCD and extracted the total events for each observation in the 0.6--1.3 keV energy range, excluding the resolved X-ray, calibration, and anomalous sources mentioned in Section~\ref{sec:obs}. The 0.6--1.3 keV range was chosen based on simulations of the expected group halo emission. Then, mean SBs for both the group and background pointings were calculated. Since the group emission is extended and much larger than the point-spread functions of the XISs, the net SB is just the subtraction of the two. For uncertainties, we purely considered Poisson noise. There was no detection of any group emission from this crude analysis. This result is not unexpected for such low temperature and diffuse objects. In our case, the data are much closer to the detection threshold. Hence, a more meticulous analysis is required by utilizing the full spectrum. Spectral analysis enables us to better constrain the background components, which then facilitates extraction of the source emission. For example, the hard energy band allows us to better pin down the active galactic nuclei (AGN; power-law) component. Therefore, we shifted our approach to precisely analyzing the spectra of each group and background pointing.

\subsection{Spectral Analysis} \label{sec:spec}
\subsubsection{Modeling Groundwork} \label{ssec:modg}

Spectra of each observation were generated using \verb+Xselect+ and we binned all spectra with a minimum of 25 photons in each bin using the FTOOL \verb+GRPPHA+. 
The instrumental response was simulated by generating redistribution matrix files using the XIS response generator \verb+xisrmfgen+ ver.\ 2012 April 21, which includes information concerning the quantum efficiency of the detectors \citep{ishi07}. Next, we used the Monte Carlo ray-tracing algorithm \verb+xissimarfgen+ ver.\ 2010 November 5 to produce the ancilliary response files (ARFs), which account for the effective area of each detector. The input GTI files were from the cleaned event files with the COR2 condition applied. 

Furthermore, we approximated emission caused by cosmic-ray and $\gamma$-ray interactions with the telescope's interior by generating the NXB spectra using the tool \verb+xisnxbgen+ ver.\ 2010 August 22, which uses the night-Earth data collected by \suz\ \citep{miz04,yama06}. Night-Earth data were accumulated for more than 750 ks for the BI CCD and 1.5 Ms for the FI CCDs, combined. Since XIS0 and XIS3 are both FI CCDs, we were able to combine their spectra, NXB, and response files using \verb+addascaspec+. To avoid systematic uncertainties in the background calibration, all spectra were fit in the energy ranges 0.6--7 keV for FI CCDs and 0.5--5 keV for the BI CCD \citep{ichi13}.
All spectra were modeled with \verb+Xspec+ \citep{arn96} ver.\ 12.8.0. Also, both the background and group observations' FI and BI spectra were fit simultaneously to improve the constraints on model parameters, since concurrently fitting all spectra maximizes S/N.  

\subsubsection{Extragalactic, Galactic, and NXB Modeling} \label{ssec:cxbmod}

The background constituents in both the outskirts and background observations were modeled by several components: NXB, Galactic emission, unresolved extragalactic sources (cosmic X-ray background, CXB), and emission due to solar wind charge exchange \citep[SWCX;][]{fuji07}. The NXB component was subtracted from each spectrum using the pregenerated NXB spectra discussed in Section~\ref{ssec:modg}. To address any possible shortcomings in the NXB generated by \verb+xisnxbgen+, we visually inspected the NXB-subtracted binned and unbinned spectra for any significant NXB excess and added Gaussian lines to model any residual NXB emission lines. 
All line normalizations were allowed to fit freely during the spectral fits. 

Unresolved extragalactic sources, i.e., AGNs, were modeled using a power-law (\emph{pow}) component with photon index ($\Gamma$) frozen at 1.41 \citep{hump06}. We accounted for Galactic gas halo emission with one absorbed \emph{apec} thermal plasma model, where the temperature was allowed to be free. 
Since NGC~5129 is close to the North Polar Spur, which is a section of the Galaxy that has enhanced X-ray emission, we added a second Galactic \emph{apec} component at $k_BT = 0.4$~keV \citep{gas07, mill08, sun09}. 
We used zero redshift and solar abundances for both background \emph{apec} models, where the temperature of the $0.4$~keV model was fixed during the spectral analysis. 
Also, the Galactic and extragalactic components were modified by a \emph{wabs} multiplicative model component to include photoelectric absorption by the Galaxy \citep{dic90}. Neutral hydrogen column densities were computed using the default parameters on the web-based $N_H$ tool,\footnote[6]{http://heasarc.gsfc.nasa.gov/cgi-bin/Tools/w3nh/w3nh.pl} where we chose the Dickey and Lockman weighted average values. Thus, the NXB-subtracted background model is \emph{wabs*(pow + apec[free] + apec[0.4]}(for NGC 5129 only)\emph{) + gau}(residual NXB lines), where the normalizations for all model components were treated as free parameters (Table~\ref{tab:bkgdpar}).

We performed simultaneous fits between the FI and BI spectra, since the Galactic, extragalactic, and galaxy group emission should correspond between different CCDs. However, the residual NXB line normalizations were allowed to fit independently due to the variation of this type of emission between differing CCDs, as well as in time. 

\subsubsection{SWCX Modeling}

SWCX \citep{fuji07} provides additional CXB to the spectra. It occurs when rapidly moving, highly ionized solar wind interacts with more neutral gas (usually hydrogen) in the solar system and strips an electron. Then, this electron enters an excited state in the solar wind ion and cascades down, releasing an X-ray. This can occur for many ions, including carbon, oxygen, and neon \citep{crav09}. Similar to residual NXB emission lines, SWCX can be modeled with Gaussian lines.
Thus, we also visually inspected the spectra for any residual lines that could be a result of SWCX. Unlike the NXB, which should not be Doppler shifted or broadened, SWCX can be due to the velocity of the solar wind. Hence, any lines that were centered within several eV of a common SWCX line were added to the model. For each observation, any emission lines with fit normalizations below $10^{-5}$~photons cm$^{-2}$ s$^{-1}$ or that had uncertainties greater than $100\%$ were removed from the model, and the model was refit.

Table~\ref{tab:lines} depicts the emission lines that were kept in the fits and their NXB and SWCX candidates. The NXB line, Au M$\alpha$, should be centered on 2.123 keV. However, we find a strong 2.195 and 2.155 keV line for NGC 3402 background BI and NGC 5129 background FI observations, respectively. \citet{sek15} also identifies lines at similar energies to this instrumental NXB line. The Suzaku Data Reduction Guide discusses this feature as a result of an improper calibration of that NXB line. We have effectively removed this calibration issue by including the residual lines in our models.

\subsubsection{Group Halo Emission Modeling} \label{ssec:grmod}

We modeled the group halo emission using an \emph{apec} thermal plasma model modified by Galactic absorption, allowing the temperatures and normalizations to vary freely and with the remaining parameters frozen at $Z = 0.2 Z_{\odot}$ (based on the measurements by \citet{eck11}) and the respective redshift of each group's central galaxy ($z = 0.0153$ for NGC~3402 and $z = 0.0230$ for NGC~5129). Here we have used the default abundance table for this version of \verb+Xspec+, \emph{angr} \citep{and89}. This group emission was added to all the background components to model the target group spectra. Then, both background and group FI and BI spectra were simultaneously fit, totaling four spectra for NGC 3402 and six for NGC 5129. There were six for NGC 5129 owing to both target group pointings and one background pointing.

Furthermore, we considered the possibility of systematic uncertainties in the background spectral modeling. To do this, we fit all combinations of models where $\Gamma$ = 1.41 or 1.56, the Galactic foreground \emph{apec} temperature would be one single component and allowed to vary or frozen at two components ($k_BT = 0.07$ keV and $0.2$ keV), and residual NXB and SWCX lines would be included or not considered entirely. The $\Gamma=1.56$ variation assumes that the power-law component is due to unresolved low-mass X-ray binaries (LMXBs) within our Galaxy, whereas the two Galactic \emph{apec} components of $k_BT = 0.07$ keV and $k_BT = 0.2$ keV reference the model parameters in \citet{hump11,hump12}. For NGC 5129, we kept the additional fixed $k_BT = 0.4$ keV \emph{apec} component for all models. 

All eight of the different background model parameters were applied to the source group and background spectra in this way, producing eight separate simultaneous spectral models. The task \emph{steppar} was performed on each fit for the group \emph{apec} temperature to ensure that it had not fallen into a local minimum. The subsequent fits were all relatively good for NGC 3402, with reasonable fit parameter values and indicating clear detections of the group halo emission. Unfortunately, for NGC 5129, the group \emph{apec} temperatures for each pointing were vastly different, where $k_BT$ was unphysically large for the 1st observation and the fits were fairly poor.

To gauge what was occurring, more spectral analysis was performed for NGC 5129. First, we tried separately fitting each observation simultaneously with the background. The eight fit results for the 1st observation were well behaved, in which the reduced $\chi^2$ and parameter values with errors were all acceptable. On the other hand, the fits for the 2nd observation were comparatively poor in reduced $\chi^2$ and some produced unrealistically low group \emph{apec} temperatures. Also, there were very large uncertainties in the group \emph{apec} normalization and $k_BT$, as well as in several other parameters. Moreover, the background components did not agree between the separate fittings of each observation.

Next, we attempted fitting the background observations separately and then constrained the target observations' best-fit background parameters to be within 3$\sigma$ of the results from those separate fits. The outcome was similar to the other test: the 1st observation had good fits, with most parameters being well constrained, whereas the 2nd observation consistently had extremely poor fits with large uncertainties in several parameters, especially the group \emph{apec} normalization and $k_BT$. The 2nd observation also generated unrealistic group temperatures in this test. From this, it is apparent that the 2nd observation of NGC 5129 is a nondetection. Though unfortunate, this is not completely unexpected due to the excessive number of extraction regions needed to filter out several point sources for that observation, as seen in Fig~\ref{fig:obs}. The fully extracted observed area was considerably smaller than that of the 1st pointing, which can be detrimental to such low-SB observations.

Thus, we chose to present the results of the simultaneous fit between the background and off-center target pointing for NGC 5129 1st. Any subsequent results referred to as ``NGC 5129" are solely from that observation. The resulting eight models each for NGC 3402 and NGC 5129 were all comparatively good fits, varying little in reduced $\chi^2$ (see Table~\ref{tab:grouppar}). 

\subsubsection{Spectral Analysis Results}
We chose the models shown in Tables~\ref{tab:bkgdpar} and~\ref{tab:grouppar} as a result of their overall excellent fit to the data, including consideration of residuals, being the models nearest to mean and median across the spectra for both groups (when distributed by temperature), and allowing the Galactic \emph{apec} temperature to vary freely. This type of model is more likely to be physically accurate, since the Galactic emission varies throughout the sky not only in normalization but also in temperature \citep{mill12}. The average best-fit temperatures for the Galactic foreground were $k_BT_{ave} = 0.177$ keV and $0.173$ keV for NGC 3402 and NGC 5129, respectively. These temperatures are consistent with those of other research using the same approach, e.g., \citet{bau09} and \citet{sim17}.
In addition, choosing the AGN index of 1.41 is far more representative of the power-law component than that of LMXBs, since our observations are relatively high in the Galactic plane with Galactic latitudes of $\sim$ 40$\degree$ and 75$\degree$ for NGC 3402 and NGC 5129, respectively. For each chosen fit, the reduced $\chi^2$ (defined as $\chi_{min}^2/dof$) is approximately unity, suggesting that the background has been successfully modeled and extracted from the hot halo group emission.

\begin{figure}[h]
	\includegraphics[width=0.35\textwidth,angle=-90]{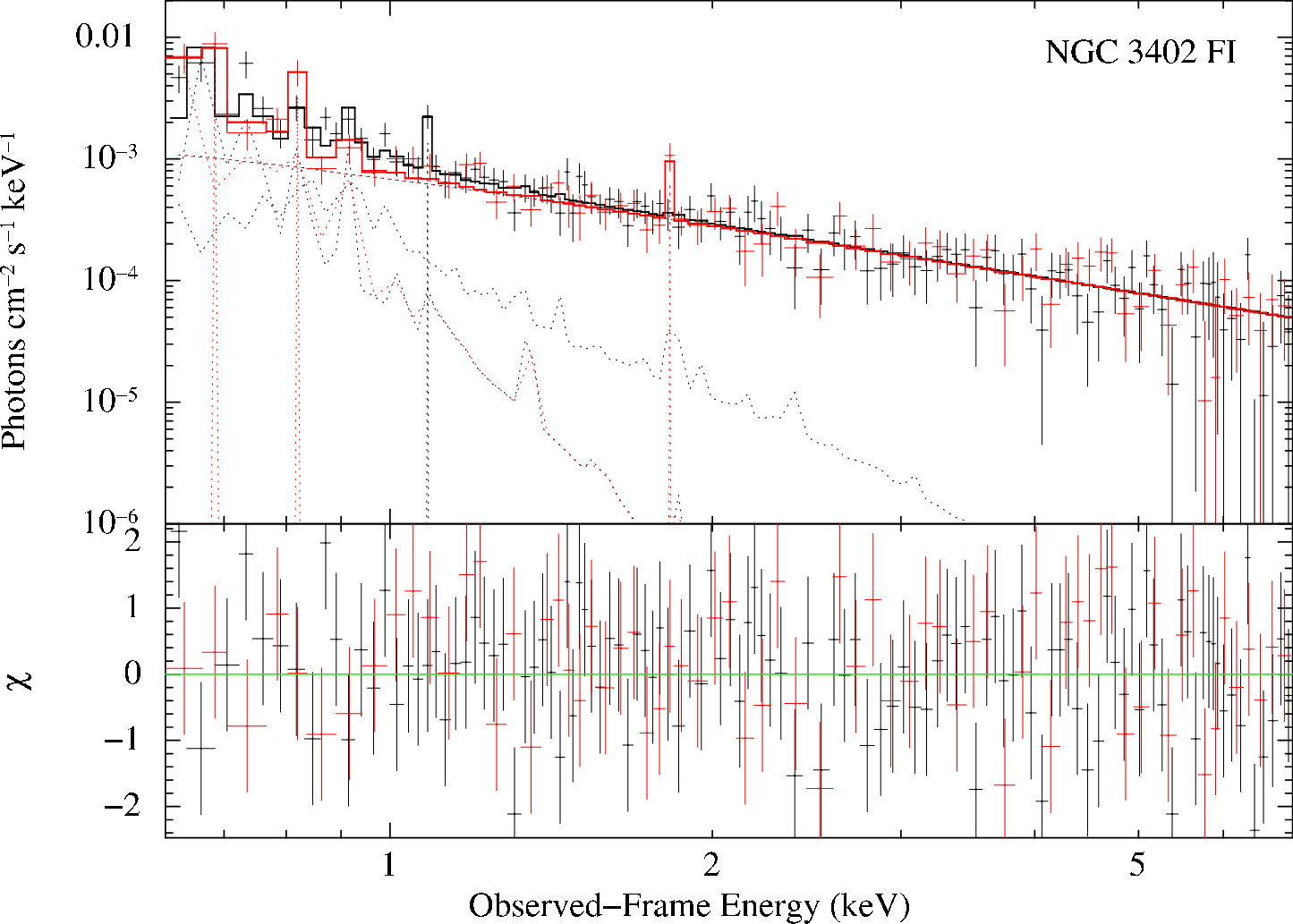}
	\includegraphics[width=0.35\textwidth,angle=-90]{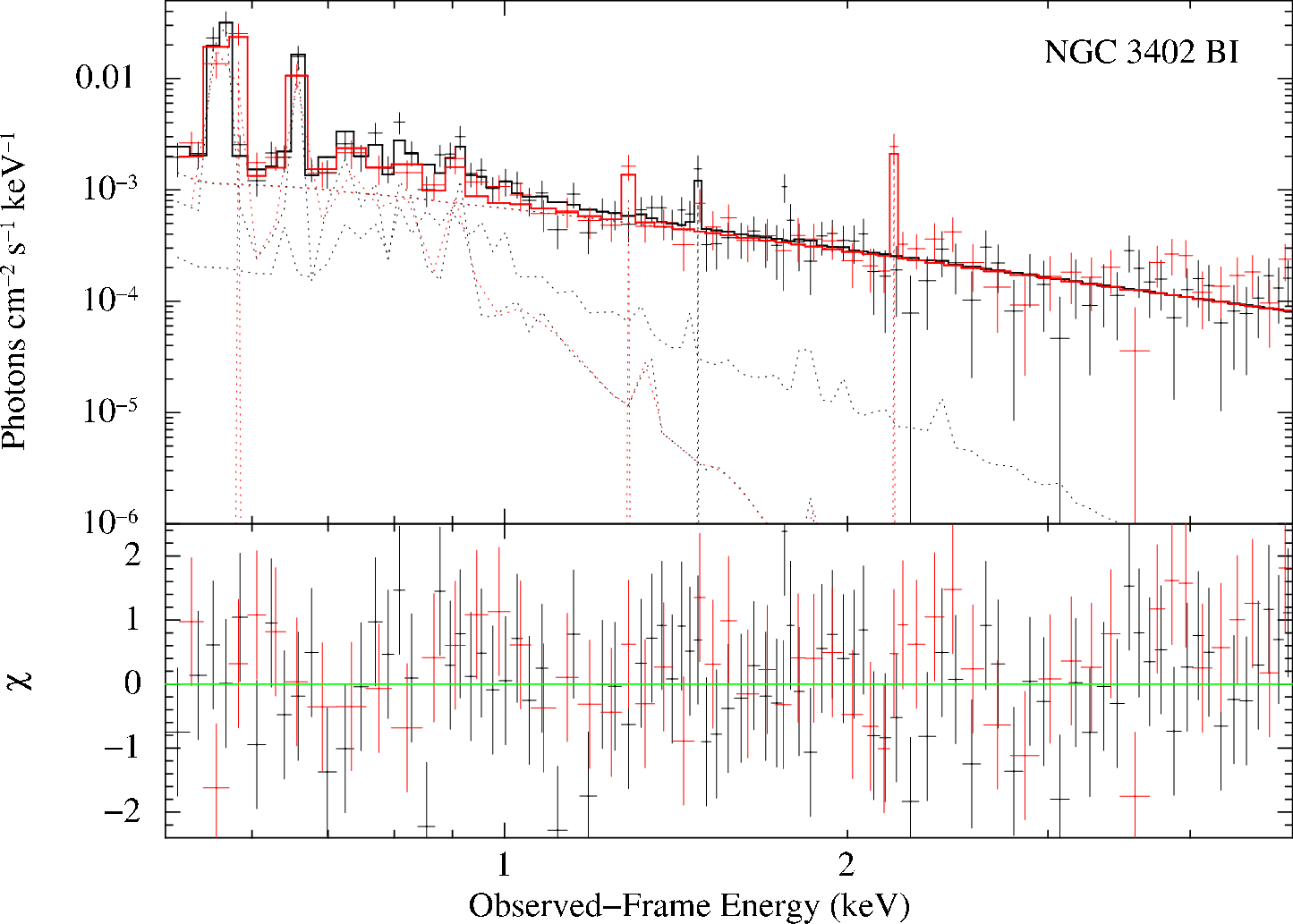}
	\includegraphics[width=0.35\textwidth,angle=-90]{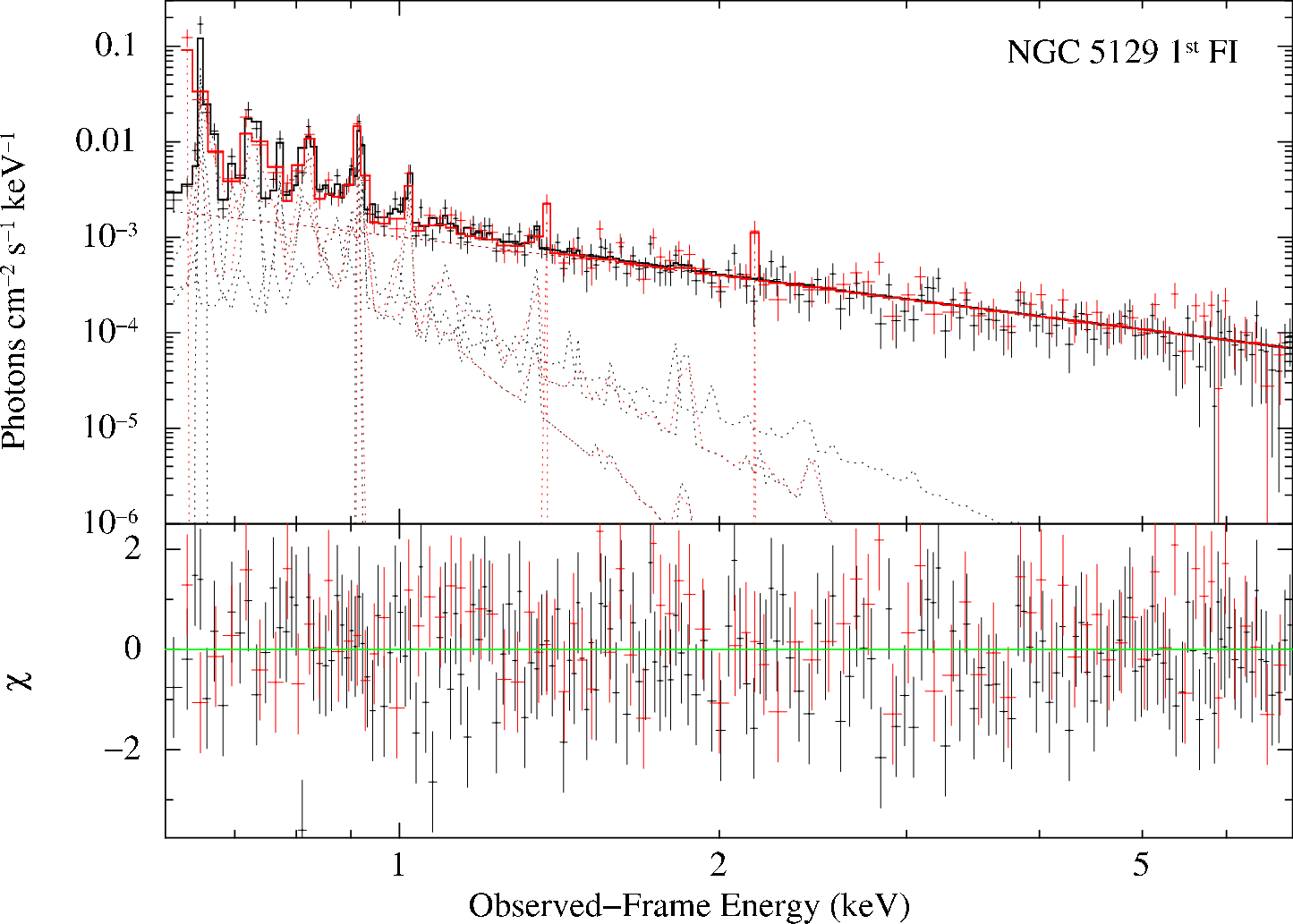}
	\includegraphics[width=0.35\textwidth,angle=-90]{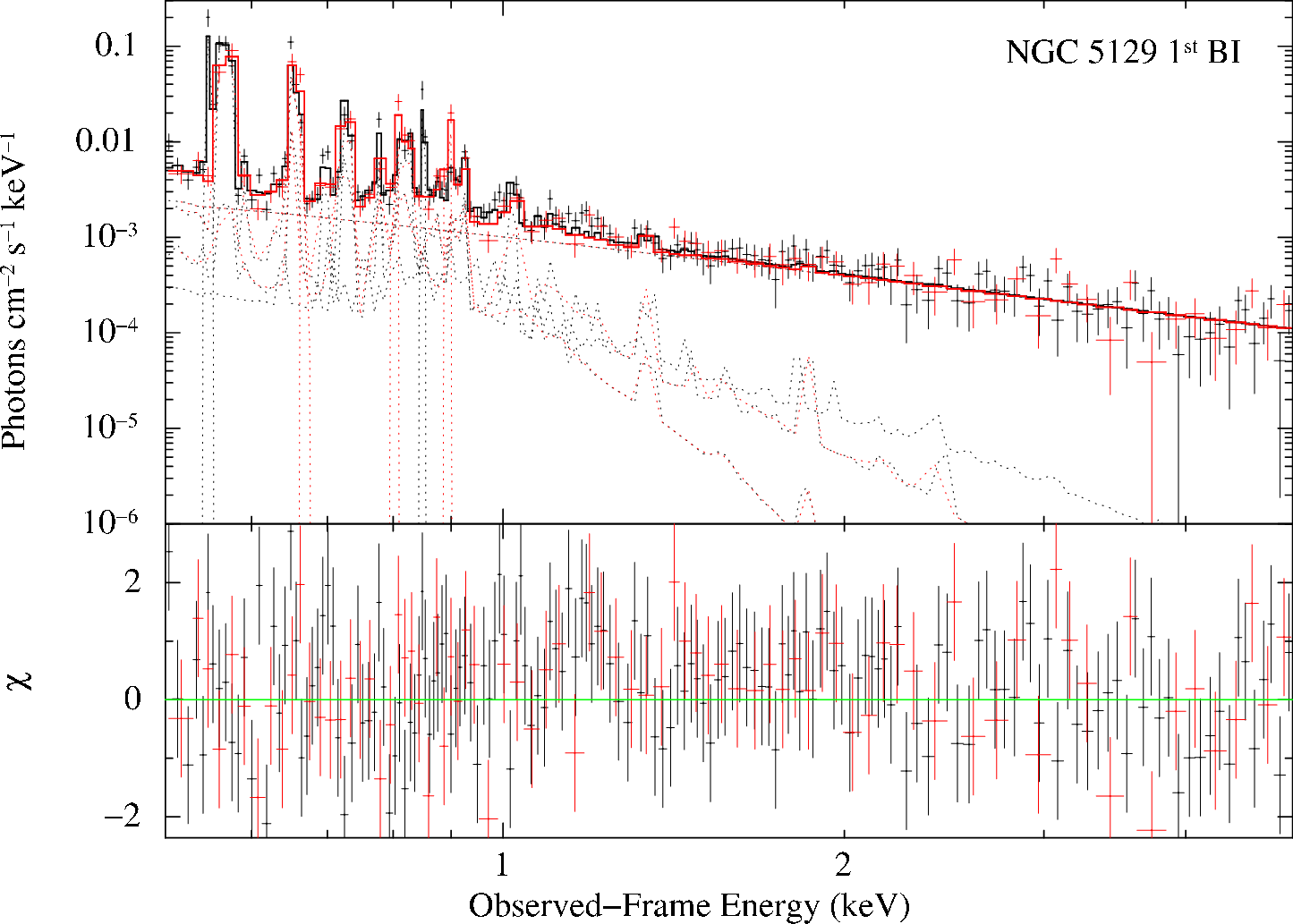}
	
	\caption{Unfolded spectra for NGC 3402 and NGC 5129 1st off-center target (black) and background (red) observations, separated into FI and BI CCDs. The solid lines are the best-fit theoretical model, not folded with the instrument response, while plus signs are the corresponding binned spectral data. Dotted lines are individual model components, where the dotted black line highest in normalization is the group halo emission. The lower panel for each spectrum contains the residuals in units of standard deviation with error bars of 1$\sigma$.}
	\label{fig:fibkobsunf}
\end{figure}

Figure~\ref{fig:fibkobsunf} shows the best-fit unfolded models (data and models deconvolved with the detector's response) with individual model components and spectral data overlaid for both the off-center group and background observations, divided into each type of CCD. These models' best-fit parameters and normalizations are given in Tables~\ref{tab:bkgdpar} and~\ref{tab:grouppar}, in which NXB and SWCX emission-line parameters and normalizations are left out for compactness.

One issue of critical importance is that when generating ARF files of extended sources, the usual assumption for \verb+xissimarfgen+ is that the emission originates from a 20$^\prime$-radius circle, so the emitting area is 20$^2\pi$ arcmin$^2$. Any normalizations generated from the fit analysis are using this area. Typically one desires the normalization to represent the extracted observation area instead. To do this, the output normalization should be multiplied by the SOURCE\_RATIO\_REG factor found in the header of the ARF file, which is simply the ratio of observation area (with extraction regions removed) to 400$\pi$. This must be considered for all analyses utilizing these normalizations or fluxes derived from them (E. Miller, private communication); see \citet{ishi07} for details. 

Also included in Table~\ref{tab:grouppar} are systematic uncertainties ($\sigma_{syst}$) in the group \emph{apec} temperature and normalization introduced from the eight variations in background parameters. While the $\sigma_{syst}$ in the group temperature is approximately half (or less) that of statistical, the $\sigma_{syst}$ in normalization is more significant. Furthermore, we changed the fixed abundance from $Z = 0.2 Z_{\odot}$ to $Z = 0.33 Z_{\odot}$ solely in the chosen models and refit. The change in group temperature between models with these abundances is small, $\Delta k_BT = 0.003$ keV for NGC 3402 and $\Delta k_BT = 0.019$ keV for NGC 5129. However, the relative change in group normalization between them is larger, $\Delta$norm/norm = 0.32 and 0.31 for NGC 3402 and NGC 5129, respectively. 

Ultimately, we chose to perform all subsequent analysis and computations purely considering statistical uncertainties. In addition, the uncertainties in group temperature and normalization shown in Table~\ref{tab:grouppar} were averaged when performing ensuing calculations. Since different components in the target spectra have been effectively isolated through spectral modeling, this allowed for more success in detecting the group emission compared to the direct subtraction method. Specifically, the group emission was detected at 4.3$\sigma$ and 2.7$\sigma$ for NGC~3402 and NGC~5129, respectively. We use these constraints on the group emission from the spectral modeling in our subsequent analysis.

\subsubsection{Stray Light} \label{ssec:stray}

Systematic uncertainties due to stray light entering the detector from the group core are crucial to consider for our off-center observations. Stray light is known as emission that scatters off the primary and secondary mirrors onto the focal plane any way other than originally intended (see the \suz\ TD and~\citealt{tak12}). For example, light outside the detector FOV can enter the mirrors and glance off the secondary mirror only. Another common scenario is radiation reflecting off the backside of the primary and then reflecting normally onto the CCD \citep{mori05}. To determine whether stray light is an important contribution for our observations, we performed simulations using the FTOOL \verb+xissim+ ver.\ 2010 November 5. We obtained a ``zeroth-order" estimate (E. Miller, private communication) by modeling both groups as point sources and used the exposure time for the cleaned events with the COR2 condition applied. The group core fluxes used in the simulations were from the total Chandra count rates (0.7--2 keV) within 1$'$, integrated from the SB profiles (Figure~\ref{fig:sbpro}). Then, we converted these to absorbed fluxes between 0.6 and 7 keV to use as inputs for \verb+xissim+. The simulations performed ray-tracing for all photons at a single energy, 1 keV, and were computed for the XIS0 CCD.

The resultant events proved to be few, with 5 counts for NGC 3402 and 1 count for NGC 5129. To estimate its significance, we compared the stray light flux (effective area approximated as 200 cm\textsuperscript{2} at 1 keV) to the 0.6--7 keV flux derived from the best-fit spectral models. We find that stray light emission constitutes less than 0.3\% when compared to the total flux from each observation, including background emission. More importantly, it comprises less than 2.8\% and 0.5\% of the absorbed group emission for NGC 3402 and NGC 5129, respectively. From this, we are confident that stray light does not play a significant role in our observations.

\subsubsection{Galactic and Extragalactic Background Variance} \label{ssec:cxbvar}

Another source of systematic uncertainty in the background modeling involves variance in the Galactic and extragalactic backgrounds. That is, these components vary spatially between the background and group observations. Variations in the X-ray emission from our Galaxy are of more concern than that of extragalactic sources. This is because extragalactic sources become more important in the hard energy band, which is why the more energetic part of the spectrum is crucial to pinning down the power-law component. Therefore, we performed additional analysis regarding the Galactic component. Moreover, we carried out analysis similar to the methods in Section 3.4 of \citet{bau09} to check the extragalactic background flux based on prior measurements in the literature.

First, we addressed the Galactic background variance by testing whether there were significant spatial variations in the Galactic component between background and group observations, where we allowed the Galactic apec temperature ($k_BT_{Gal}$) and normalization ($k_{Gal}$) to be different between the group and background observations. This means that the model was no longer simultaneously fitting the Galactic component between group and background observations. However, we simultaneously fit the NPS \emph{apec} model for NGC 5129 as in the original model.

Comparing the resulting Galactic \emph{apec} temperatures and normalizations within these unconstrained models, we found that they agree well within 1$\sigma$ uncertainties for both galaxy groups. Furthermore, those values are consistent within 1$\sigma$ of the best-fit parameters from the original models. Together, this strongly implies that there is no significant spatial variation in the Galactic background between our group and background observations. Perhaps more importantly, the best-fit group \emph{apec} emission temperatures and normalizations were also consistent and within 1$\sigma$ of each other between the original and unconstrained models.

Another helpful comparison to make involves the Galactic flux from the spectral model and the corresponding ROSAT flux. \citet{sun09} performed this for the \chan\ spectral results of their galaxy groups, which included NGC 3402 and NGC 5129. Their Figure 2 illustrates the consistency between their Galactic flux density in the ROSAT band (0.47--1.21 keV) and the ROSAT flux associated with each galaxy group's location. Similarly, we found that the Galactic flux densities using our \suz\ spectral analysis results and our values are consistent with those from \citet{sun09}, considering uncertainties.

As for the extragalactic variance, we calculated the expected background SBs for unresolved point sources (B) using AGN source count measurements (log$N$--log$S$ from \citealt{mor03}, M03). Our results are $B_{M03}$ = 6.12 $\times~10^{-12}$ (0.5--2 keV) \flux~deg$^{-2}$ and 1.55 $\times~10^{-11}$ \flux~deg$^{-2}$ (2--8 keV), where we assumed the same limiting flux $S_{excl}$ = $10S_{14}$ as \citet{bau09} for our \suz\ observations, since our exposure times and those in \citet{bau09} are similar and we do not have XMM-Newton observations to lower that limit.

In addition, we used the more recent relations and parameters in \citet{dai15} (D15, Equation (3) and Swift-all from their Table 11) and obtained $B_{D15}$ = 3.57 $\times~10^{-12}$ \flux~deg$^{-2}$ (0.5--2 keV) and 7.04 $\times~10^{-12}$ \flux~deg$^{-2}$ (2--8 keV). We compare these values to those derived from the flux of the power-law component in our spectral model fits. Our results lie between (4.44--6.57) $\times~10^{-12}$ \flux~deg$^{-2}$ (0.5--2 keV) and (1.13--1.56) $\times~10^{-11}$ \flux~deg$^{-2}$ (2--8 keV) for both galaxy groups. These values correspond extremely well with those we obtained using the parameters and equations in \citet{mor03}. They are also consistent within at most a little over a factor of two of the values computed from \citet{dai15}. Therefore, we can see that the power-law component was properly modeled in our work and we did not undersubtract the extragalactic background.

\section{Radial Profiles} \label{sec:rad}
\subsection{AtomDB} \label{sec:atom}

The release of the atomic database AtomDB ver.\ 2.0.2 in 2011 caused significant changes in the derived spectral properties of plasma with $k_BT < 2$ keV, due to updates in the Fe L-shell data \citep{fos12}. The major quantity affected for our analysis is the gas temperature, which increased by 10\%--20\% from ver.\ 1.3--2.0 and later versions. This is important to consider, since we are adding our contributions using ver.\ 2.0.2 to data derived from older AtomDB versions. To estimate the temperature change in the inner profile, we compared the projected \chan\ temperature profile of NGC 3402 from \citet{sun09} (which used AtomDB ver.\ 1.3.1) to the \chan\ data reprocessed with CIAO 4.6.1 and CALDB 4.6.2 (post AtomDB ver.\ 2.0.2; E. O'Sullivan, private communication). By determining the vertical shift between temperature profiles and averaging them, we found that the temperature measurements increased by 18.6\% between pre-AtomDB 2.0.2 and post-2.0.2 analyses. This shift was applied to the subsequent temperature and entropy profiles of the inner data, as well as the global temperatures for these objects, as mentioned in the Introduction. These adjusted temperatures are used repeatedly in our analysis and are indicated as such in the corresponding figures and text.

\subsection{Emission-weighted Radius} \label{sec:emission}

Since results from the spectral analysis are weighed by emission, we also computed the corresponding radii for each target observation. These emission-weighted radii ($R_{emw}$) were calculated by summing over all distances between each pixel in the extraction region and the group's X-ray center, multiplied by the SBs at those pixel locations. As discussed below, these SBs were estimated using a model of the SB profile at the outskirts. Then, we divided by the sum of the SBs at those radii, producing a radius that correlates to the emission-weighted center of each observation.

To obtain the SB function for the outskirts of each group, we used data from \chan\ observations \citep{sun09}, as well as the SBs determined from our spectral analysis, adjusted for the \chan \ detector and energy band (see \S\ref{sec:surfb}). First, we selected outskirts data such that the cutoff corresponded to the innermost extent of our \suz \ observations without extraction regions applied: 200 and 90 kpc for NGC 3402 and NGC 5129, respectively. Next, we fit a power law to these outer data, allowing the normalization and power-law index to be free. These data include the SBs obtained in this work at the initial central location of each observation.
From this, we weighed all radii according to the summations mentioned previously and iterated this process until the radii reached convergence. The subsequent emission-weighted radii were  $R_{emw}$ = 375 kpc for NGC 3402 and 249 kpc for NGC 5129. This corresponds to an average mass overdensity of $\Delta$ = 530 and 1430 times the critical density of the universe, for NGC 3402 and NGC 5129, respectively.

\begin{figure}[h!]
	\includegraphics[width=.5\textwidth]{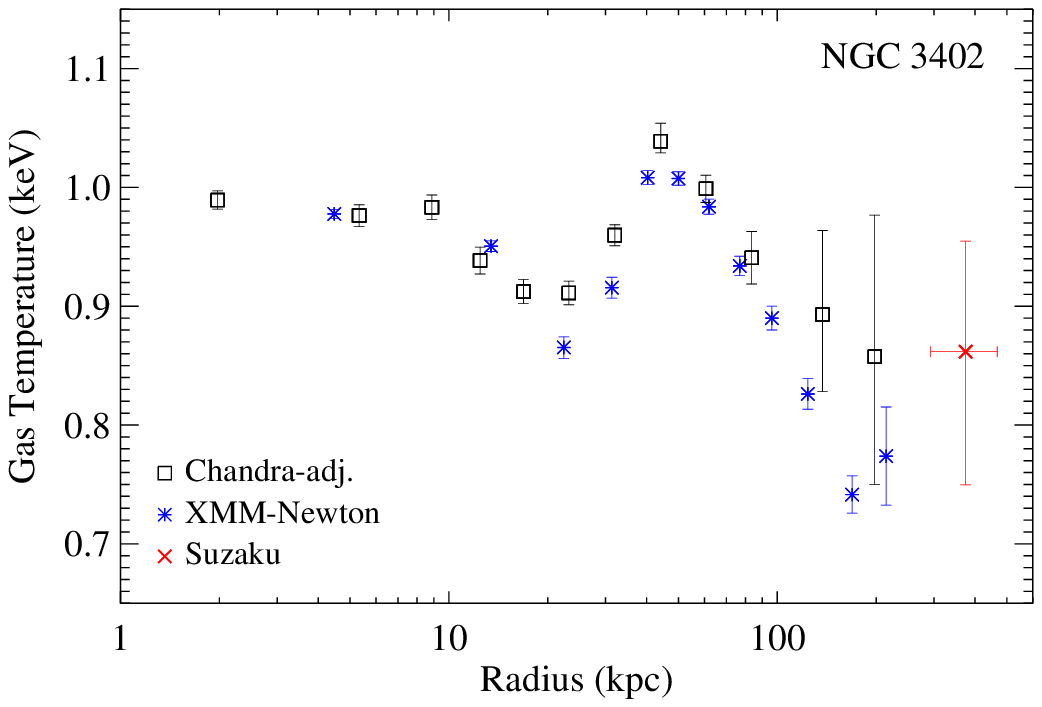}
	\includegraphics[width=.5\textwidth]{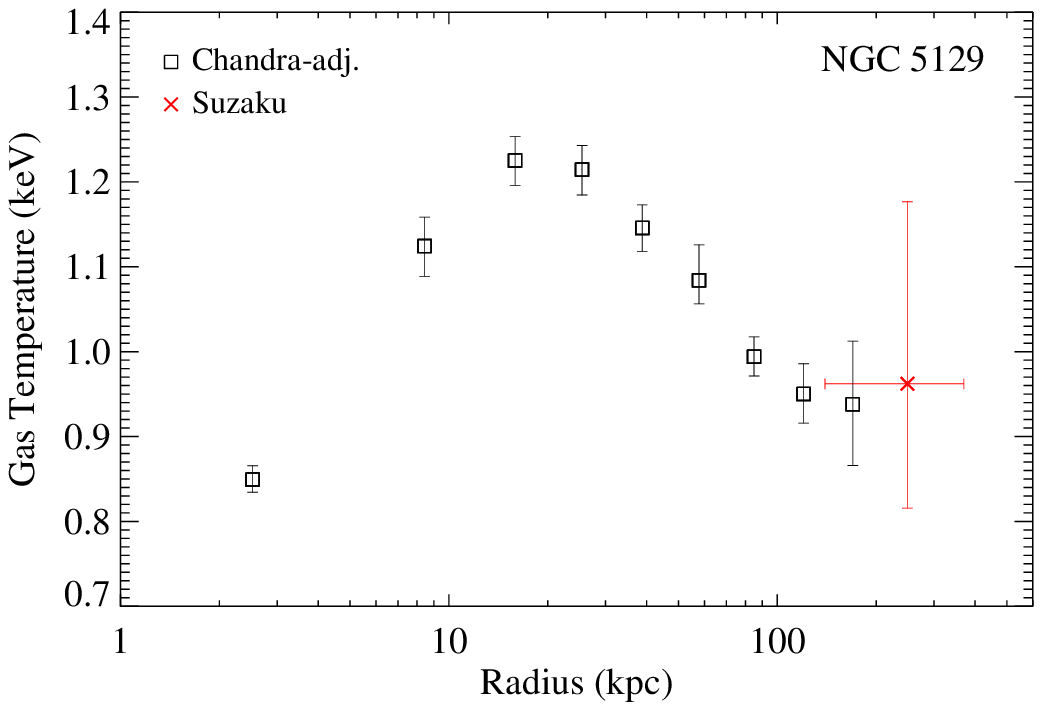}
	\caption{Projected temperature profiles with 1$\sigma$ uncertainties in $k_BT$ and emission-weighted radial bin sizes overlaid. Black squares are \chan\ data retrieved from \citet{sun09} and adjusted to AtomDB ver. 2.0.2, blue asterisks are projected XMM-Newton data (E. O'Sullivan, private communication), and red crosses are our \suz\ values.}
	\label{fig:temppro}
\end{figure}

\begin{figure}[h!]
	\includegraphics[width=.5\textwidth]{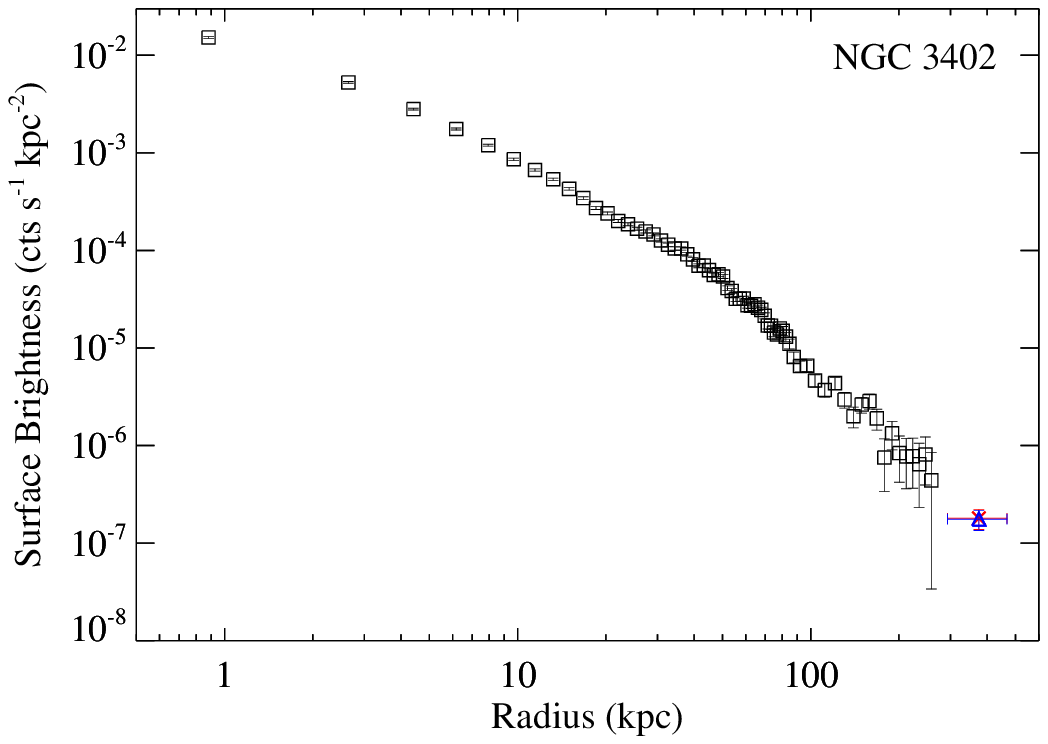}
	\includegraphics[width=.5\textwidth]{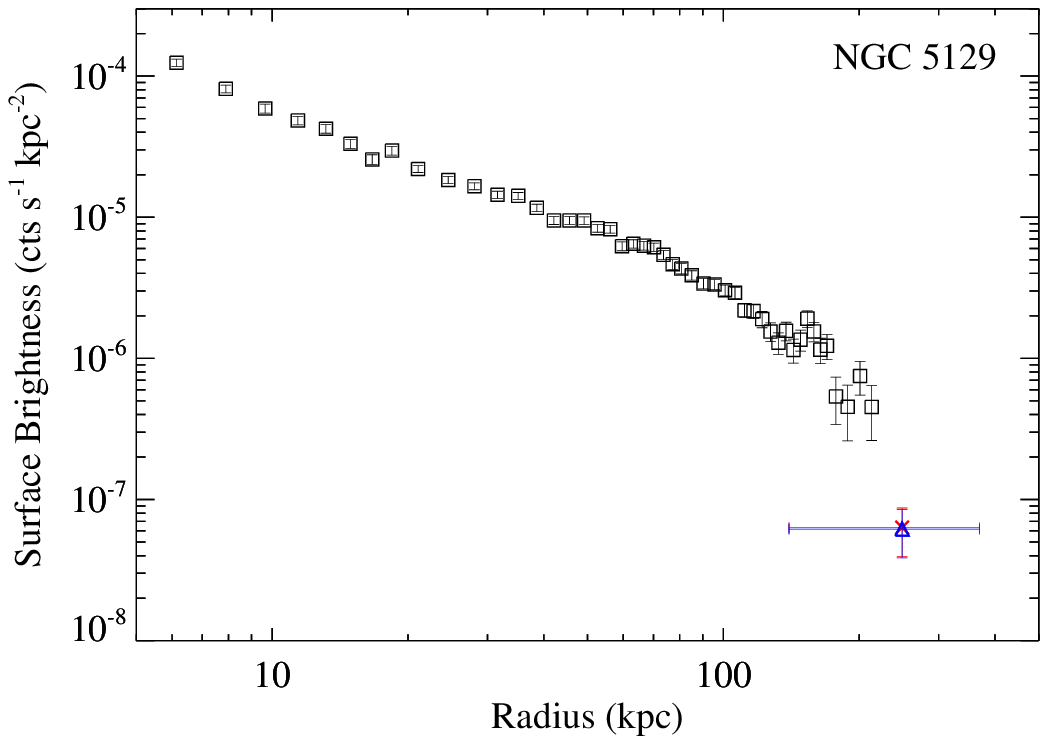}
	\caption{Projected SB profiles with 1$\sigma$ uncertainties in SB and emission-weighted radial bin sizes overlaid. Black squares are \chan\ data, while red crosses and blue triangles are our \suz\ FI and BI data, respectively. Note that the SB has not been divided by the effective area of the telescope, which is energy dependent.}
	\label{fig:sbpro}
\end{figure}

Lastly, using the iterated outskirts SB function, we computed a radial bin size based on the locations within which 68\% of the total halo emission for each observation is contained, centered on $R_{emw}$. Specifically, we found the radius at which 16\% of all emission within the extraction region was contained and set this as the lower bound. The upper bound was found using the corresponding location within which 84\% of emission was enclosed. These bin sizes have been overlaid for all radial profiles (Figures~\ref{fig:temppro},~\ref{fig:sbpro},~\ref{fig:2beta},~\ref{fig:calib},~and~\ref{fig:entropy}).

\subsection{Gas Temperature} \label{sec:temp}

Figure~\ref{fig:temppro} illustrates the projected temperature profiles for NGC 3402 and NGC 5129 out to $R_{530}$ and $R_{1430}$, respectively, by combining our outer \suz\ data with the inner, adjusted \chan\ data \citep{sun09}.  
Here we have plotted the asymmetric uncertainties originally found through \verb+Xspec+ instead of the symmetrized ones used in all other related calculations. 
Furthermore, we included the projected temperature profile derived from \xmm\ observations of NGC 3402 using AtomDB ver. 3.0 and SAS 13.5 (E. O'Sullivan, private communication), based on the work by \citet{osul07}.

Comparing the \chan\ and \xmm\ profiles of NGC~3402, we can see an overall agreement between them, where both temperature profiles exhibit ``wiggles'' that match in radii.
\citet{osul07} discussed the temperature dip at $\sim$ 10--40 kpc as the possible presence of a ``cool core that has been partially re-heated by AGN activity,'' resulting in a region of warmer gas enclosed within a shell of cool gas. They also discuss the possibility of the shell being due to a recent merger. \citet{lagan19} support the latter hypothesis given the nature of their 2D spectral maps of NGC 3402. Their metallicity map shows a clear increase along the southwest-to-northeast direction in the region of that shell, which they deem can only be the result of merging activity. Moreover, both the \chan\ and \xmm\ data show decreases in temperature at $R > 50$~kpc. 
The \suz\ emission-weighted temperature at 375~kpc, $k_BT = 0.86\,\pm\,0.10$~keV, is consistent with the outermost \chan\ and \xmm\ data points. Our contribution appears to indicate a leveling off of the outer profile as opposed to decreasing, which may also be the result of a merger or perhaps shock heating of infalling material. Yet this should not be overemphasized owing to the large uncertainties involved. 

In the case of NGC 5129, our \suz\ temperature measurement from the 1st observation is also comparable to the outermost \chan\ data points, albeit slightly larger.  
However, considering the relatively large uncertainties, it is in agreement with the declining trend of the inner data, typical of a universal temperature profile (e.g., \citealt{vik05}). In this way, NGC 5129 is far closer to exhibiting the general shape of relaxed galaxy clusters' and groups' temperature profiles than NGC 3402.

\subsection{Surface Brightness} \label{sec:surfb}

The projected SB profiles in Figure~\ref{fig:sbpro} were produced by combining our \suz\ measurements at $R_{emw}$ (Table~\ref{tab:sbnd}) with inner data from the \chan\ observations in \citet{sun09}. 
We have converted the \suz\ count rates (CRs) into \chan\ ACIS-S 0.7--2.0 keV CRs (same as in \citealt{sun09}) using \verb+WebPIMMS+. Our \suz\ data expand on measurements of the SB profiles, especially in the case of NGC~3402, in which the profile is extended by $\sim$117~kpc. 
As expected, our SB measurements are lower than the inner SBs and fall on the declining trends established by the inner data.

Total CRs of the two groups were measured to greater than 0.62\,$R_{500}$, by interpolating and integrating the SB profiles. Combining this with the adjusted global temperatures for each group, $k_B \overline{T}_{3402} = 0.88$~keV and $k_B \overline{T}_{5129} = 0.90$~keV \citep{sun09}, the estimated 0.5--2 keV unabsorbed X-ray fluxes are $F_{X,3402} = (9.09 \pm 0.20) \times10^{-12}$~\flux\ and $F_{X,5129} = (1.790~\pm~0.042) \times10^{-12}$~\flux. We determined bolometric luminosities of $L_{Xbol,3402} = (7.00~\pm~0.15) \times10^{42}$~\lumin\  and $L_{Xbol,5129} = (3.157~\pm~0.074) \times10^{42}$~\lumin. Also, we extrapolated the bolometric luminosities out to $R_{500}$ and $R_{200}$. Furthermore, we found the X-ray luminosities in the ROSAT 0.1--2.4 keV band to be $L_{ROSAT,3402} = (6.77 \pm 0.15) \times10^{42}$~\lumin\ and $L_{ROSAT,5129} = (3.042 \pm 0.071) \times10^{42}$~\lumin. The aforementioned values can be seen in Table~\ref{tab:fbmass}.

\subsection{Electron Number Density} \label{sec:nden}

Following \citet{hud10} and \citet{eck11}, the X-ray SB in units of photons~s$^{-1}$~cm$^{-2}$~arcsec$^{-2}$ at some projected distance on the sky ($R$) can be expressed in terms of the emission measure along the line of sight, $EM = \int n_en_H \mathrm{d}l$, by 
\be \label{eq:sbint}
\Sigma = \int_{-\infty}^\infty \! n_en_H \, \mathrm{d}l \  \frac{\int_{E_1}^{E_2}\Lambda(T,Z,E)\mathrm{d}E}{4\pi (1+z)^4},
\ee 
where $\Lambda(T,Z,E)$ is the ``emissivity function for a plasma of temperature $T$ and metallicity $Z$ at energy $E$'' \citep{hud10}, $z$ is the galaxy group redshift, $n_e$ is the electron number density, and $n_H$ is the number density of hydrogen.
Converting to deprojected, three-dimensional radius $r$ and assuming $n_e \approx 1.2 n_H$,
since the ratio of the number of H to He is approximately 10\% and most electrons come from H and He in these systems \citep{arn05}, Equation~\ref{eq:sbint} becomes
\be \label{eq:sbndint} 
\Sigma(R) = 2\int_R^\infty \! \frac{n_e^2(r)}{1.2} \, \frac{r\mathrm{d}r}{\sqrt{r^2 - R^2}} \frac{\int_{E_1}^{E_2}\Lambda(T,Z,E)\mathrm{d}E}{4\pi (1+z)^4}.
\ee 
Here the \emph{apec} normalization is defined as
\be \label{eq:apecnorm}
k \equiv \frac{10^{-14}}{4\pi [D_A(1 + z)]^2}\int n_e n_H \mathrm{d}V, \ee   
where $D_A$ is the angular diameter distance, which can be found using the group redshift.
To calculate $n_e(r)$ from Equation~\ref{eq:sbndint}, we needed to measure the projected SB $\Sigma(R)$ and determine its shape in order to find the final shape of the $n_e(r)$ profile. As discussed below, $\Sigma(R)$ was estimated using data from another work and the $\Sigma$ at $R_{emw}$ from our \suz\ observations through additional spectral analysis. To do this, we first utilized the inner $n_e$ data produced from \chan\ observations (\citealt{sun09}) to pin down the type of modeling needed to fit the SB profile including our \suz\ data.

\begin{figure}[h!]
	\includegraphics[width=0.5\textwidth]{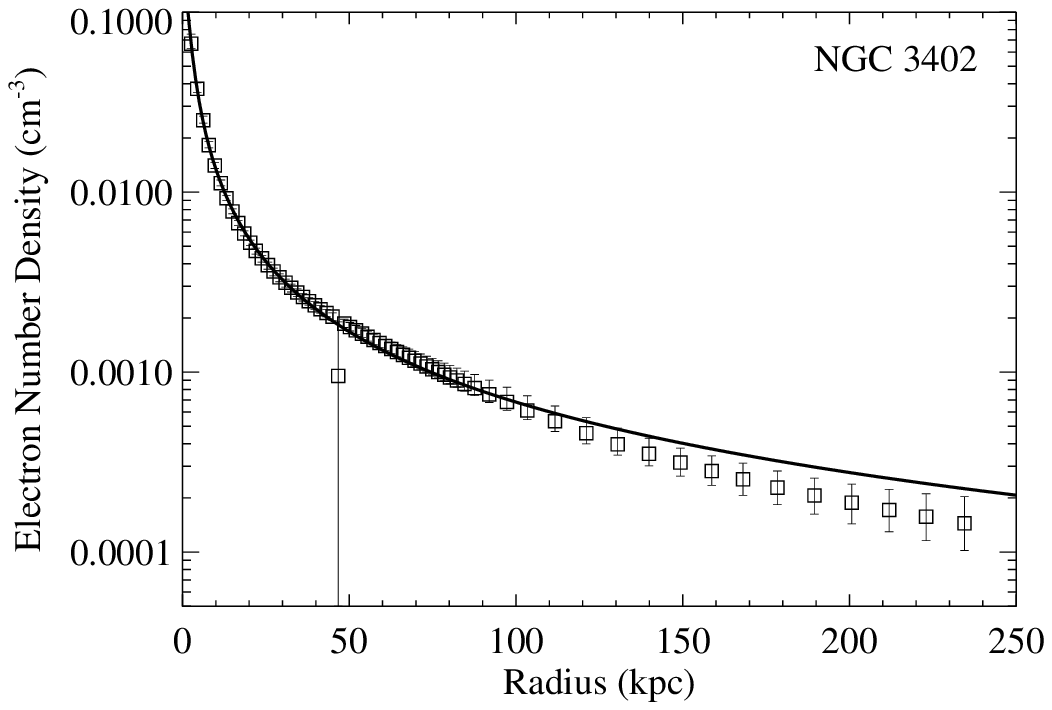}
	\includegraphics[width=0.5\textwidth]{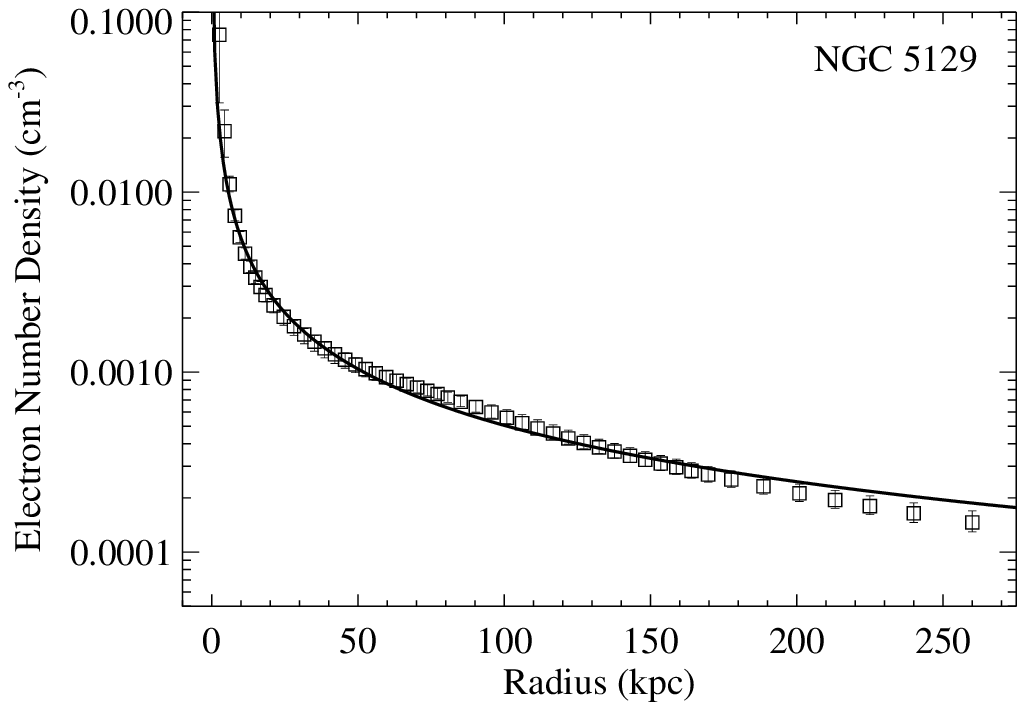}
	\caption{Single $\beta$-model fits (black lines) to the deprojected \chan\ data from \citet{sun09}.}
	\label{fig:1beta}
\end{figure}

\begin{figure}[h!]
	\includegraphics[width=0.5\textwidth]{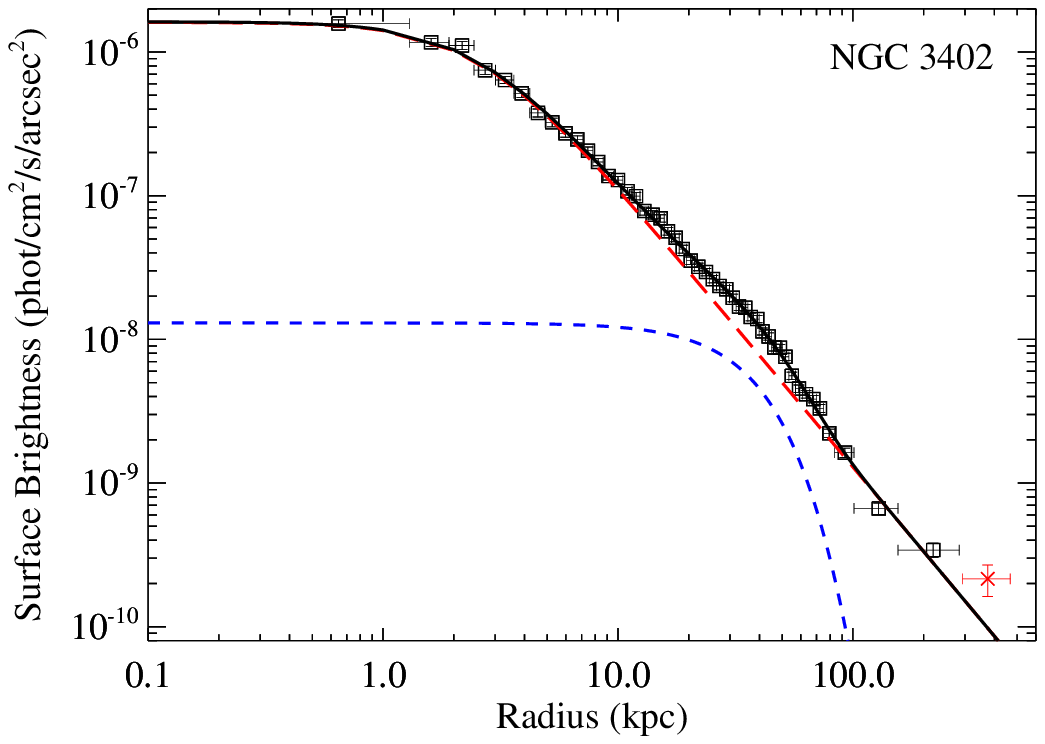}
	\includegraphics[width=0.5\textwidth]{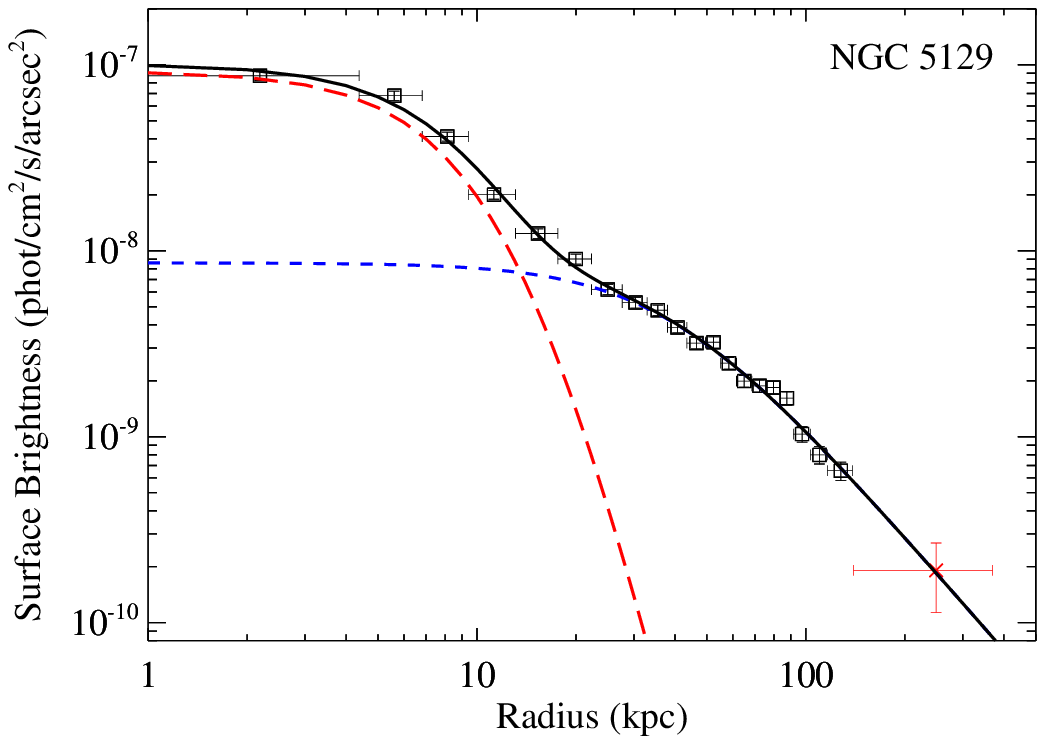}
	\caption{2$\beta$-model fits to the projected \chan\ data (black squares) from \citet{eck11} plus our \suz\ data (red crosses). Red and blue dashed lines are the first and second $\beta$-model components, respectively, while black is the sum of the two.}
	\label{fig:2beta}
\end{figure}

Most galaxy clusters' and groups' X-ray number densities and SBs can be well described by the class of models called $\beta$-models (\citealt{breg07} and references therein), which assume that the gas is isothermal and in hydrostatic equilibrium.
In a single $\beta$-model assuming spherical symmetry, the electron number density of the gas is parameterized by
\be \label{eq:ndbeta} n_e(r) = n_{e0}\left(1+ \left(\frac{r}{r_c}\right)^2\right)^{-\frac{3\beta}{2}}, 
\ee 
where $n_{e0}$ is the central electron number density (the value of $n_{e}$ at $r = 0$), $r_{c}$ is the core radius, and $\beta$ is the slope of the density profile, typically observed to be $\sim$ 0.5 for groups \citep{mul00}. Thus, Equations~\ref{eq:sbndint} and \ref{eq:ndbeta} reduce to
\be \Sigma(R) = \Sigma_0\left(1 + \left(\frac{R}{r_c}\right)^2\right)^{-3\beta + \frac{1}{2}},
\ee 
where $\Sigma_0$ is the SB at $R = 0$.
This single $\beta$-model form is sufficient for many rich clusters, but it is overall a poor fit to the emission from groups \citep{mul00}. To test this, we began with the single $\beta$-model and fit the \chan\ number density data for each group obtained by \citet{sun09}. 
Though initially asymmetric, we symmetrized the uncertainties in the \chan\ data. Unless otherwise stated, all uncertainties used in the calculation of subsequent quantities and their errors have been symmetrized. Figure~\ref{fig:1beta} shows that the single $\beta$-model is indeed not a good fit to the group data, especially at large radii where our \suz\ observations take place. Therefore, since it is clear that a more complicated model is needed, we chose to use a two-component $\beta$-model, or a 2$\beta$-model. The 2$\beta$-model is characterized by the addition of two $\beta$-models, each with separate core radii, betas, and central SBs: 
\be \label{eq:sb2beta} 
\Sigma(R) = \Sigma_{01}\left(1 + \left(\frac{R}{r_{c1}}\right)^2\right)^{-3\beta_1 + \frac{1}{2}} + \Sigma_{02}\left(1 + \left(\frac{R}{r_{c2}}\right)^2\right)^{-3\beta_2 + \frac{1}{2}}.
\ee 

Now that the need of at least a 2$\beta$-model is apparent, we decided to fit 2$\beta$-models to the SB profile from \citet{eck11} (which are their Figures C.19 and C.27), along with our \suz\ data. To perform uncertainty estimation on the best-fit parameters, we fixed each parameter for which the uncertainty was being computed while letting the others vary and calculated the $\chi^2$ over a range of fixed parameter values. This procedure assumes that there is no covariance between parameters. The 1$\sigma$ uncertainty occurs when $\chi^2 = \chi^2_{min} + 1$, i.e., $\Delta\chi^2 = 1$.

However, this revealed degeneracies in parameters, especially in $r_{c2}$ and $\beta_2$ for NGC 3402 but also in $r_{c1}$ and $\beta_1$ for NGC 5129. They were severely correlated for NGC 3402, resulting in unphysically high values of one parameter, while the value of the other increased and still produced $\Delta\chi^2 < 1$. One way to alleviate this is to fix the degenerate core radius to its best-fit value. With that parameter now completely fixed, the other five were allowed to be free, the data refit, and  $\Delta\chi^2 = 1$ uncertainties approximated as initial guesses. This was done for both galaxy groups and, along with switching to brute-force uncertainty estimation, brought about reasonable fit parameters and uncertainties.

Our SB profiles visually fit all data very well for NGC 5129, whereas the \suz\ data from this work is over 1$\sigma$ above the best-fit model for NGC 3402. This may indicate that a different model (perhaps a 3$\beta$-model) would better fit the data. Results of these fits are shown in Figure~\ref{fig:2beta} and Table~\ref{tab:2betafit}, where the minimum $\chi^2$ is larger than ideal considering the degrees of freedom. Nevertheless, a fit solely to the inner data (generated by \citet{eck11}) produced an analogous $\chi^2$/dof, so we felt justified in proceeding with the analysis. Note that we assume dof = $N - P$, where $N$ is the number of data points and $P$ are the free parameters, five for these fits. However, the degrees of freedom could be as high as dof~$ = N - 1$ for nonlinear models \citep{and10}, which would improve the reduced  $\chi^2$.

\begin{figure}[h!]
	\fbox{\includegraphics[width=0.5\textwidth]{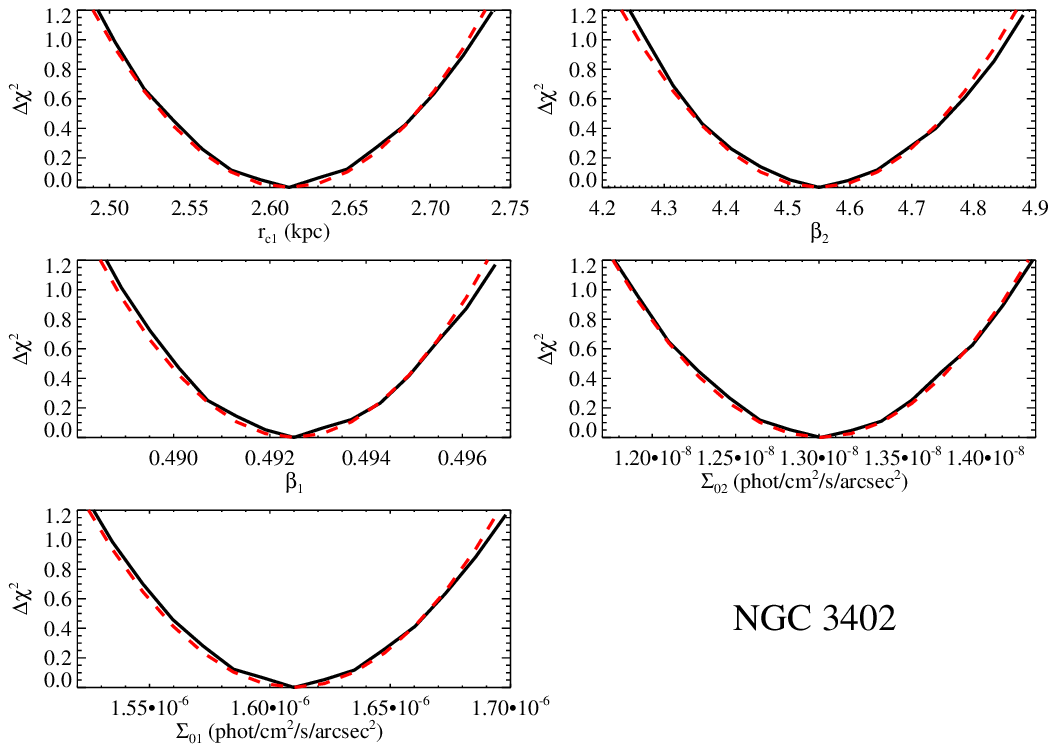}}
	\fbox{\includegraphics[width=0.5\textwidth]{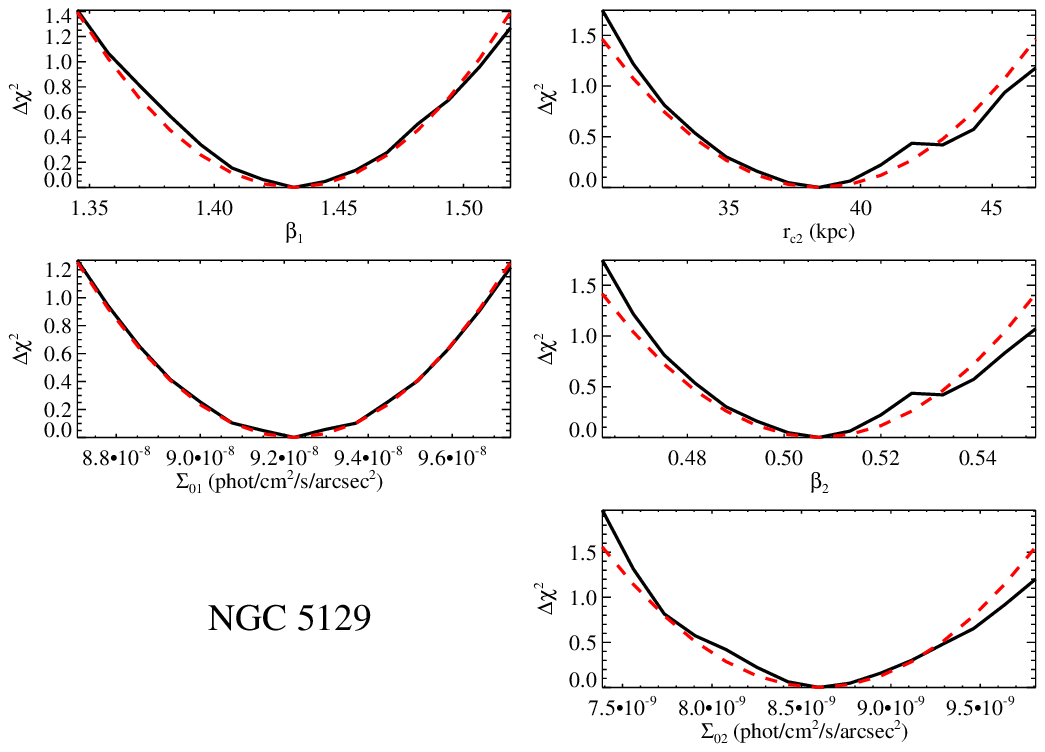}}
	\caption{Quadratic fits (red dashed lines) to the 2$\beta$-model parameters (black solid lines) represented via their $\Delta\chi^2$ value over the 5-dimensional grid, from which their 1$\sigma$ uncertainties were derived.}
	\label{fig:gridpars}
\end{figure}

\begin{figure}[h!]
	\includegraphics[width=0.5\textwidth]{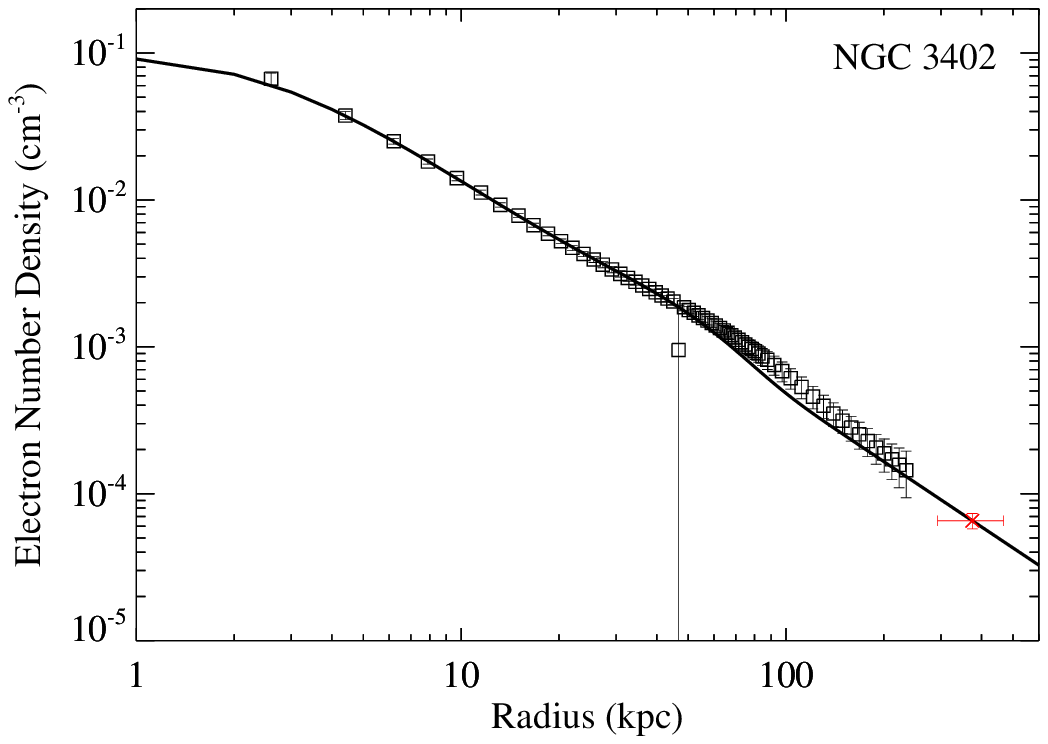}
	\includegraphics[width=0.5\textwidth]{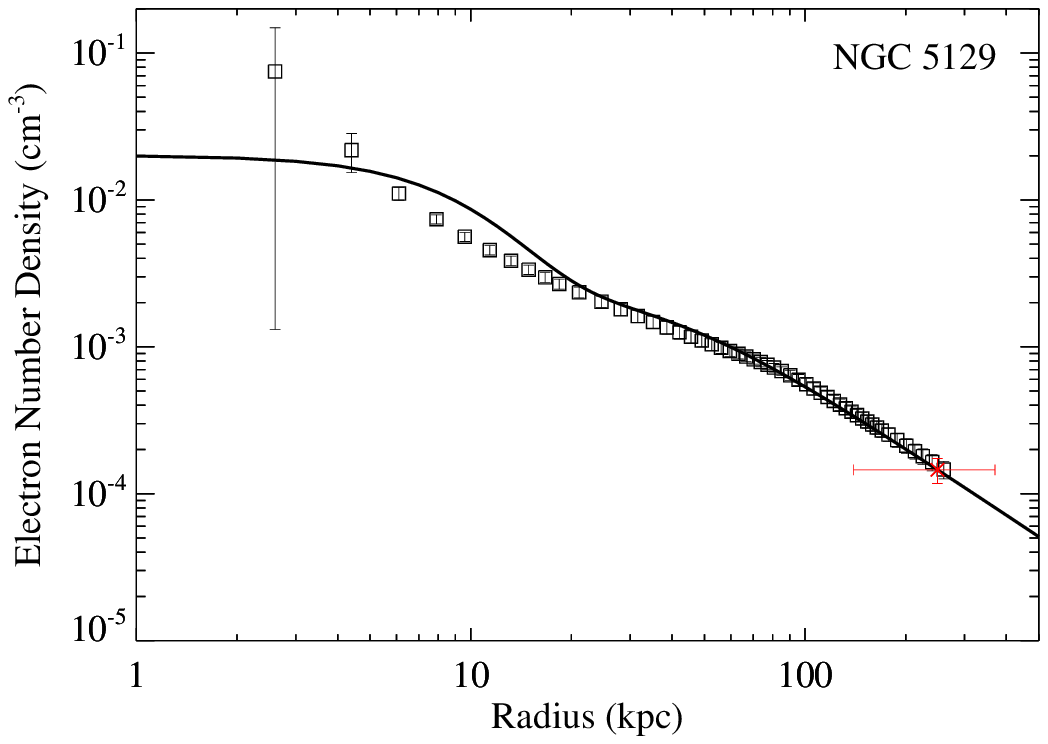}
	\caption{\citet{sun09} $n_e$ data (black squares) and our \suz\ data (red crosses), with radial bin sizes overlaid. The black line is the calibrated 2$\beta$-model profile derived from the full SB data set (\citealt{eck11} and \suz).}
	\label{fig:calib}
\end{figure}

Brute-force uncertainty estimation involves calculating $\chi^2$ over a grid of parameter ranges for a model chosen to characterize a data set. In this case, that is a 5-dimensional grid of the 2$\beta$-model parameter values, where $r_{c2}$ and $r_{c1}$ have been fixed for NGC 3402 and NGC 5129, respectively. The ranges were based on the aforementioned initial $\Delta\chi^2=1$ uncertainty estimates, in which the $r_{c}$ values had been fixed. Using 15 values in each of the five dimensions, we obtained model parameters and their 1$\sigma$ uncertainties by finding the minimum $\chi^2$ for each value across the entire grid of all other parameters. From that, the $\Delta\chi^2$ for each parameter's range of values was found by subtracting out the global minimum $\chi^2$, and then a quadratic was fit to each to determine $\Delta\chi^2=1$, i.e., the 1$\sigma$ confidence interval. Table~\ref{tab:2betafit} provides best-fit 2$\beta$-model parameters with 1$\sigma$ uncertainties found using this brute-force method. See also Figure~\ref{fig:gridpars}, which illustrates quadratic fits to the $\Delta\chi^2$ for each parameter.

Now that the SB model is determined and the parameters have uncertainties, other quantities can be derived such as the $n_e$ profile. For a 2$\beta$-model where the SB is in the form of Equation~\ref{eq:sb2beta}, the $n_e(r)$ can be written as
\be \label{eq:ne2beta} 
n_e(r) = \left[n_{e01}^2\left(1 + \left(\frac{r}{r_{c1}}\right)^2\right)^{-3\beta_1} + n_{e02}^2\left(1 + \left(\frac{r}{r_{c2}}\right)^2\right)^{-3\beta_2}\right]^{\frac{1}{2}}.
\ee 
Substituting Equation~\ref{eq:ne2beta} into Equation~\ref{eq:sbndint} and integrating yields Equation~\ref{eq:sb2beta}, where $\Sigma_{01}$ and $\Sigma_{02}$ are related to $n_{e01}$ and $n_{e02}$ by
\be \label{eq:s0iton0i}
\Sigma_{0i} \equiv n_{e0i}^2\frac{ \int_{E_1}^{E_2}\Lambda(T,Z,E)\mathrm{d}E}{4\pi (1 + z)^4 1.2} LI_i. \ee
Here $LI_i$ is the line integral defined as
\be \label{eq:lineint}
LI_i \equiv \int_{-\infty}^\infty \left(1 + \left(\frac{l}{r_{ci}}\right)^2\right)^{-3\beta_i}\mathrm{d}l.
\ee
Combining this with $\Sigma_{12} = \Sigma_{01}/\Sigma_{02}$, Equation~\ref{eq:ne2beta} becomes
\be \label{eq:nefinal}
n_e(r) = \eta\left[\Sigma_{12}LI_2\left(1 + \left(\frac{r}{r_{c1}}\right)^2\right)^{-3\beta_1}+LI_1\left(1 + \left(\frac{r}{r_{c2}}\right)^2\right)^{-3\beta_2}\right]^{\frac{1}{2}}, 
\ee
in which $\eta = n_{e0}/\sqrt{\Sigma_{12}LI_2 + LI_1}$ and $n_{e0}^2 = n_{e01}^2 + n_{e02}^2$ is the electron number density at $r = 0$. Finally, $n_{e0}$ can be determined by inserting Equation~\ref{eq:nefinal} into Equation~\ref{eq:apecnorm}, resulting in
\be \label{eq:n0def}
n_{e0}^2 = \frac{4\pi10^{14}(\Sigma_{12}LI_2 + LI_1)D_AD_L1.2k}{\Sigma_{12}LI_2EI_1 + LI_1EI_2} ,
\ee
where $D_L$ is the luminosity distance we found using the group redshift and $EI_i$ is defined as the ``emission integral divided by the central [electron] density" \citep{hud10} and is expressed as
\be \label{eq:emint}
EI_i = 2\pi\int_{-\infty}^\infty \int_{0}^R R\left(1 + \frac{R^2 + l^2}{r_{ci}^2}\right)^{-3\beta_i}\mathrm{d}R\mathrm{d}l.
\ee
Therefore, the $n_e$ profiles can be derived from the fit results of the SB profile (Table~\ref{tab:2betafit}) and the \emph{apec} normalization, $k$. These profiles have been calibrated to account for the \emph{apec} normalization derived from our \suz\ outskirts observations being applied to the entire group. Illustrated in Figure~\ref{fig:calib} are the calibrated profiles along with the $n_e$ data in Figure~\ref{fig:1beta} and our $n_e$ at $r_{emw}$. There is a deviation in the data for NGC 3402 in the range of $r \approx 60-140$ kpc that matches our profile well considering uncertainties. Similarly, there is a small deviation between $r \approx$ 5 and 18 kpc for NGC 5129. However, the data and profile match extremely well in the outskirts, which is our area of interest for this work. 

Next, we calculated 1$\sigma$ uncertainties for $n_e$ at the emission-weighted radius. Since we want to probe as much probability space as possible, we produced grids spanning parameter values to at least $\Delta\chi^2 > 6.63$, which encompasses the confidence interval out to 99\% for one parameter of interest. This was done for 25 steps in each parameter, which approached the limit of what was computationally feasible using this method. With these grids of parameter and $\chi^2$ values, we performed maximum likelihood estimation, where we computed corresponding likelihoods and obtained probabilities. This produced the probability distribution with respect to $n_e$ at $r_{emw}$. The mean values were chosen to be the $n_e$ associated with the global minimum $\chi^2$ for the full grid, and the 1$\sigma$ uncertainties were found by taking the 68\% area under the normalized probability distributions, centered on the mean. For both groups, these mean quantities were located at the peaks of the largely symmetric distributions. There is slight asymmetry in the distribution for NGC 5129, yet this is to be expected considering the parameter curves in Figure~\ref{fig:gridpars}. The resultant $n_{e}$ at $r_{emw}$ can be seen in Table~\ref{tab:sbnd}.

To obtain the total number density of the hot gas, we assumed $n\mu = n_e\mu_e$, where $\mu$ is the mean molecular weight and $\mu_e$ is the mean molecular weight per free electron. Assuming total ionization, $n_e \approx 1.2n_H$ and $\mu \approx 0.62$, $\mu_e \approx \left(X + \frac{1}{2}(Y + Z)\right)^{-1} \approx 1.18$, in which $X$ = 0.7, $Y$ = 0.29, and the metallicity is $Z$ = 0.2$Z_{\odot}$ = 0.004. 

\subsection{Entropy} \label{sec:entr}

\begin{figure}[h!]
	\includegraphics[width=.5\textwidth]{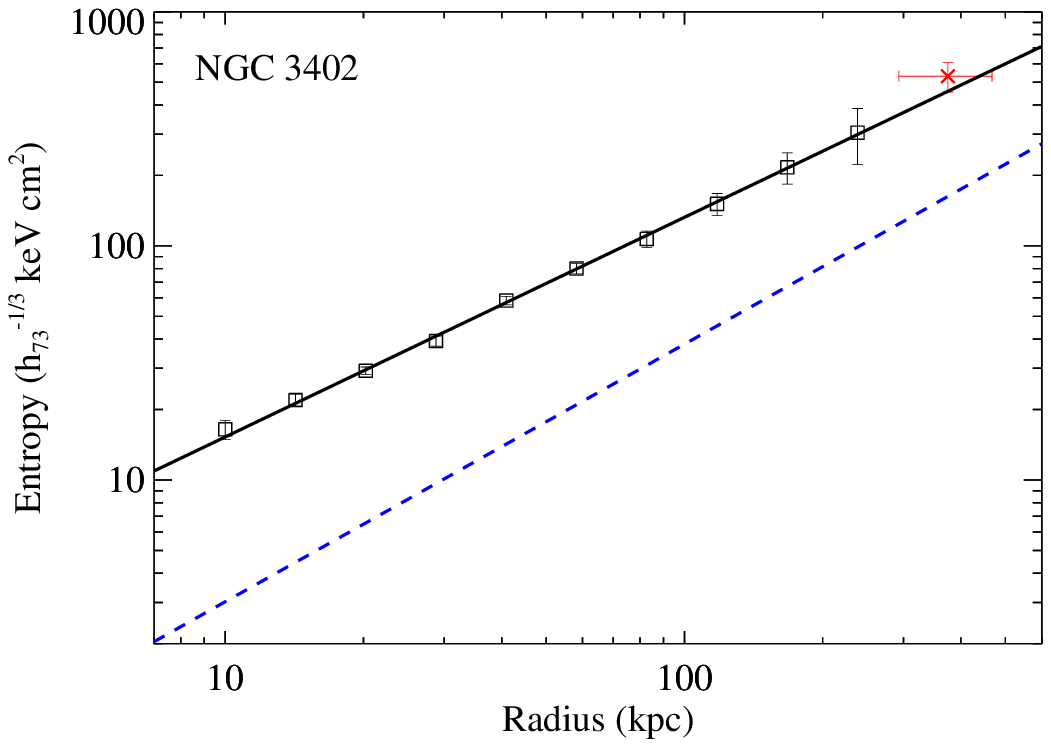}
	\includegraphics[width=.5\textwidth]{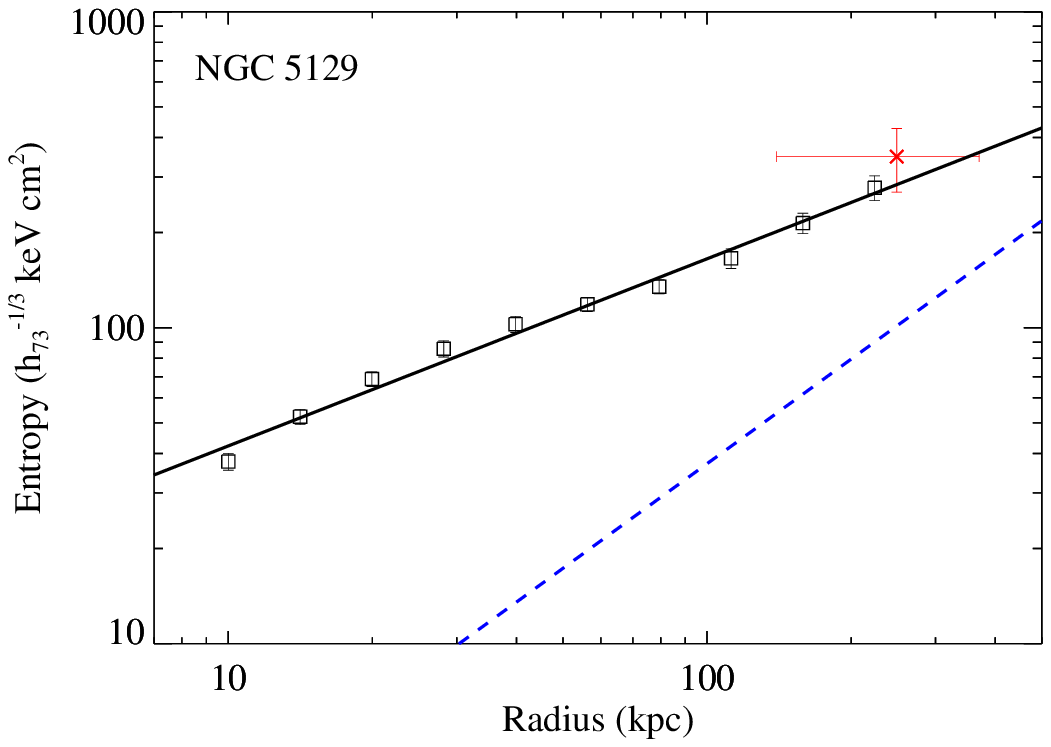}
	\caption{Entropy profiles in which black squares are Chandra data from \citet{sun09} (adjusted to the recent version of AtomDB) and red crosses are our \suz\ data with radial bin sizes overlaid. The solid black lines are power-law fits to the data, whereas the dashed blue lines are the self-similar models as discussed in Section~\ref{sec:entr}.}
	\label{fig:entropy}
\end{figure}

The entropy of the intragroup medium (IGM) is given by $K=k_BT/n_e^{2/3}$, where $k_BT$ is in keV. Technically, this is the term inside the usual thermodynamic entropy per particle equation (multiplied by a constant, $C$) for an adiabatic, monatomic gas: $\kappa = k_B\cdot\ln{CK} + \kappa_0$. Here $\kappa_0$ is described as a ``constant that depends only on fundamental constants and the mixture of particle masses" \citep{voit05}. Regardless, the former quantity is widely used and called ``entropy" because this representation can separate the effects due to gravitational and nongravitational processes. See \citet{cav09} and \citet{voit05} for more details. Figure~\ref{fig:entropy} shows the data determined from the spectral analysis of the outskirts by this work, where we have used the symmetric uncertainties in the outer $n_{e}$ and group \emph{apec} $k_BT$ to compute the uncertainty in entropy. These entropies are $K_{emw}$ = 530 $\pm$ 76 keV$ $cm$^{2}$ for NGC 3402 and $K_{emw}$ = 348 $\pm$ 79 keV$ $cm$^{2}$ for NGC 5129 (Table~\ref{tab:fbmass}).

There appears to be no tendency for the entropy in either group to drop off or flatten in the outskirts, the latter of which has been observed in clusters \citep[e.g.,][]{geo09,hos10, kawa10}. In fact, our \suz\ data indicate that the opposite may be occurring, although this finding is inconclusive considering the uncertainties and radial bin sizes in the outskirts. This tendency is more pronounced in NGC 5129, where the outer entropy appears to be significantly higher than the trend of the inner data. Furthermore, we have included in Figure~\ref{fig:entropy} the self-similar models, which are the entropy profiles solely due to gravitational processes \citep[$K\propto r^{1.1}$;][] {wong16}. We also plotted power-law fits to the data, including the contributions from this work. The best-fit power-law index, $\Gamma$, for NGC 3402 was $\Gamma$ = 0.94, whereas for NGC 5129 the index was much flatter at $\Gamma$ = 0.59.

\section{Mass Determination} \label{sec:mass}
\subsection{Hot Gaseous Halo and Stellar Masses} \label{sec:mgas}

The gas mass density can be given by $\rho_{gas}(r) = m_p\mu_en_e(r)$, where $m_{p}$ is the mass of a proton. 
Assuming spherical symmetry, we can calculate total gas masses using our 2$\beta$-model best-fit parameters out to the emission-weighted radii of our \suz\ observations. Here we applied the same grid of parameter values and method used to derive the $n_{e}$ at $r_{emw}$ in Section~\ref{sec:nden}. The gas mass was $M_{gas,3402}$ = (9.3 $\pm$ 1.1) $\times 10^{11} M_\odot$ for NGC~3402 and $M_{gas,5129}$ = (6.1 $\pm$ 1.2) $\times 10^{11} M_\odot$ for NGC 5129.

For estimating the stellar mass component of both groups, we used the Two Micron All Sky Survey $K_{s}$-band apparent magnitude of each member galaxy, since emission in the near-infrared is less affected by interstellar extinction and the stellar mass-to-light (M/L) ratios in this band vary relatively little over a large range of star formation histories \citep{bell01,bell03}. To determine the galactic membership, we implemented the SIMBAD Astronomical Database to obtain papers analyzing group membership. For NGC 5129, \citet{mah04} found 19 member galaxies out of $N_{obs}$ = 33 total galaxies in the observation field. However, NGC 3402 was unique in that there were two differing sets of galaxies considered to be possible group members: six from \citet{cro07} and four from \citet{guz09}. Two of these galaxies overlap, one being the brightest group galaxy (BGG) (NGC 3402), which resulted in eight different member candidates. Using the most current radial velocity data from each paper, we further narrowed down the membership criteria using a redshift cutoff based on the velocity dispersion of the groups. To obtain the velocity dispersion, $\sigma_{disp}$, we used the scaling relation for groups and clusters, $\sigma_{disp} = 309$ \kms\ $(k_BT/1$ keV$)^{0.64}$ \citep{xue00}, where $k_BT$ is the global temperature adjusted for AtomDB. The resulting velocity dispersions were $\sigma_{disp, 3402}$ = 285~\kms~and $\sigma_{disp, 5129}$ = 289~\kms.

Constraining each galaxy to be within twice that dispersion around its group's global radial velocity, the remaining galactic memberships were $N_{3402,czcut} = 5$ and $N_{5129,czcut}$ = 19, matching the findings in \citet{mah05} and \citet{mah04}, respectively. This method is similar to the ``sigma clipping" procedure used by \citet{mah05}. Furthermore, to be consistent with the other mass measurements in this paper, we restricted the membership criteria such that each galaxy must lie within $r_{emw}$. This resulted in $N_{3402}$ = 4 and $N_{5129}$ = 5. Also, we adopted a stellar $K_{s}$ mass-to-light ratio of $\Upsilon = 0.9$, in which a 30\% 1$\sigma$ uncertainty inferred from Figure~18 in \citet{bell03} was applied. Thus, the uncertainty of the mean is 0.3$\Upsilon/\sqrt{N}$, where $N$ is the number of member galaxies in each group.

Combining this with the $K_{s}$-band magnitudes of each member galaxy ($m_{K}$), we obtained $M_{*,3402} = (2.87 \pm 0.43) \times 10^{11} M_{\odot}$  and $M_{*,5129} = (7.11 \pm 0.96) \times 10^{11} M_{\odot}$. As expected, the stellar masses are dominated by the central elliptical galaxies, also called BGGs. Assuming a 30\% 1$\sigma$ uncertainty, these masses are $M_{*BGG,3402} = (2.23 \pm 0.67) \times 10^{11} M_{\odot}$ and $M_{*BGG,5129} = (3.8 \pm 1.1) \times 10^{11} M_{\odot}$. In addition, we extrapolated the hot gas and stellar masses for both groups out to characteristic radii, $r_{500}$ and $r_{200}$. For the stellar masses, we simply extended the distance criteria for the group member candidates out to those radii. These quantities are listed in Table~\ref{tab:fbmass}, along with other mass components and parameters.

The contribution of cold gas is considerably less than that of hot gas in these types of systems. Combining this knowledge with the large uncertainties in the other mass components, the effect of the cold molecular gas is negligible here and is not considered in our analysis. 

\subsection{Total Gravitational Mass} \label{sec:mtot}

Under the assumption of hydrostatic equilibrium, the total mass enclosed within a certain radius (in this case $r_{emw}$) is
\be \label{eq:mtot1} M_{tot}(< r_{emw}) = \frac{- k_BT(r)r^2}{G\mu m_p} \left( \frac{\mathrm{d}ln\rho_g(r)}{\mathrm{d}r} + \frac{\mathrm{d}lnT(r)}{\mathrm{d}r}\right), 
\ee 
where $G$ is the gravitational constant and the best-fit 2$\beta$-models are used in the first term.
Assuming isothermality, the second term in Equation~\ref{eq:mtot1} is eliminated and the $k_BT(r)$ in the first term is replaced with the adjusted global temperature given in the Introduction. By observing the temperature profiles, one can see that assuming isothermality is acceptable for NGC 3402. However, this assumption is not valid for NGC 5129, where it resembles a universal temperature profile for galaxy clusters. Therefore, we utilized a profile from \citet{sun09}, specifically their Equation (5), 
\be 
\label{eq:tprof} \frac{T}{T_{2500}} = (1.22\,\pm\,0.02) - (0.79\,\pm\,0.04)\frac{r}{r_{500}}, 
\ee 
where $T_{2500}$ is the projected temperature with its uncertainty at $R_{2500}$ for NGC 5129, which was obtained from Table 3 in \citet{sun09} and adjusted for the change in AtomDB. Thus, after applying the 2$\beta$-models and $T(r)$, Equation~\ref{eq:mtot1} becomes
\be \label{eq:mtot2} 
M_{tot}(< r_{emw}) = \frac{3k_BT(r)r^3}{G\mu m_p} \left( \frac{\Sigma_{12}LI_2\xi_1 + LI_1\xi_2}{\Sigma_{12}LI_2\zeta_1 + LI_1\zeta_2} - \frac{s}{T(r)}\right), 
\ee
where $s$ is the derivative of Equation~\ref{eq:tprof} with respect to $r$ and
\ba \label{eq:xizeta} 
\zeta_i = \left(1 + \left(\frac{r}{r_{ci}}\right)^2\right)^{-3\beta_i} \text{and}~~  \xi_i = \frac{\beta_i}{r_{ci}^2}\left(1 + \left(\frac{r}{r_{ci}}\right)^2\right)^{-3\beta_i - 1}.
\ea

Using the same grid method and parameters to obtain $n_{e}(r_{emw})$ and $M_{gas}$, we obtained total dynamical masses within the emission-weighted radii of $M_{tot,3402} = (1.750~\pm~0.013) \times 10^{13} M_{\odot}$ and $M_{tot,5129} = (1.39~\pm~0.12) \times 10^{13} M_{\odot}$, which are typical values for poor groups. Furthermore, we computed the total enclosed masses out to $r_{500}$ and  $r_{200}$, as done for many other derived properties (Table~\ref{tab:fbmass}).

\section{Discussion} \label{sec:disc}

Using \suz\ observations of two low-temperature poor galaxy groups, NGC 3402 and NGC 5129, we measured a range of properties for these groups out to $r_{530}$ and $r_{1430}$, respectively. These properties include the SB, flux, temperature, electron number density, entropy, gravitational mass, and gas halo mass. Thus, we have added NGC 3402 to the small sample of poor groups with well-measured X-ray properties out to approximately \rfive.

\begin{figure}[h!]
	\centering
	\includegraphics[width=.75\textwidth]{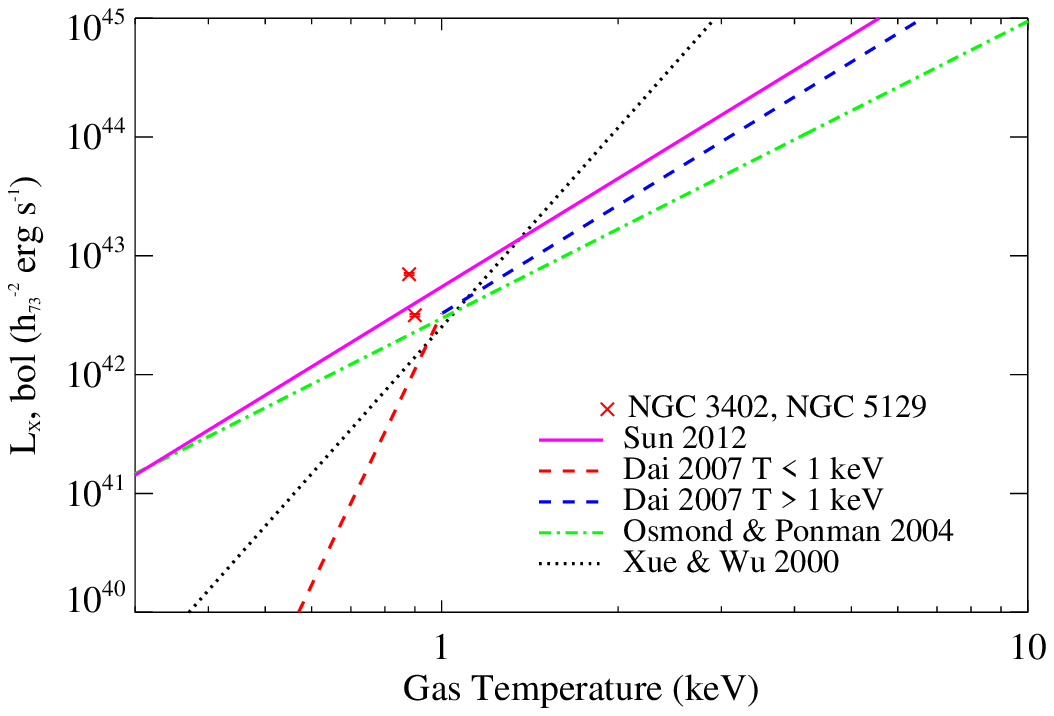}
	\caption{Bolometric X-ray luminosity (0.1--100 keV) plotted vs. global gas temperature for NGC 3402 and NGC 5129, along with their 1$\sigma$ uncertainties. Also plotted are various $L_{X}$--$T$ relations from the literature, corrected for our cosmology.}
	\label{fig:lxvst}
\end{figure}

One area of interest lies in how the $L_X$--$T$ relation differs between galaxy groups and clusters. We first compare the bolometric X-ray luminosities determined in this work for our groups and their global group temperatures to the $L_X$--$T$ relations of other works in Figure~\ref{fig:lxvst}. Plotted are our data and the relations from \citet{xue00}, \citet{osm04}, \citet{dai07}, and \citet{sun12}, in which we have adjusted the relations to our cosmology. We plotted the Poisson model fit for \citet{dai07} and the group fits for the remaining $L_X$--$T$ relations. The chosen relations were fit based on limited data from galaxy groups and thus vary widely in slope and normalization. 

Our data agree best with the shallow sloped relations by \citet{sun12} and \citet{osm04}, showing no breaks in the $L_X$--$T$ relation down to temperatures of 0.9~keV. 
Therefore, X-ray-selected (bright) clusters and groups may show universal scaling relations without breaks.
Accurate measurements for even lower temperature groups are needed to test whether the $L_X$--$T$ relation breaks at $T \lesssim 0.8$~keV. 
The optically selected groups, i.e., \citet{dai07}, have X-ray luminosities below the $L_X$--$T$ relations established from the X-ray-selected groups (all other relations in Figure~\ref{fig:lxvst}). This was independently measured in the group regime by \citet{and15}.

As for the entropy profiles, one can see that the profile for NGC 3402 lies at nearly a constant value above the $r^{1.1}$ self-similar model \citep{wong16}, representing the entropy purely due to gravitational processes. On the other hand, the profile for NGC 5129 appears to rapidly converge with the self-similar model at large radii. Clearly, the effect of nongravitational processes dominates in these systems, in the central regions of both groups and beyond for NGC 3402. That is to be expected if NGC 3402 has indeed recently undergone a merger (which would contribute to stellar feedback) or reheating due to an AGN outflow near the core. Increased energy injection not due to gravitational effects has most likely occurred in the outskirts of this group as well. Possible causes for this excess entropy could be AGN feedback that has reached the outer radii, which has been seen in other groups \citep{sun09}. As for NGC 5129, the offset in the central regions could also be a product of strong merging or AGN activity, which has radiated vast amounts of energy over its evolution. These results are expected for the significantly weaker gravitational potential wells of poor galaxy groups. Furthermore, the fact that our entropy contributions from this work both lie above the fits instead of flattening out suggests additional key differences between the group and cluster regimes. 

Another property of extreme interest is the baryon fraction. Through normal propagation of errors and assuming no covariances, we combined measurements of the gas, stellar, and gravitational masses and obtained the baryon fractions, $f_b = (M_* + M_{gas})/M_{tot}$. Measured out to $r_{emw}$, we found $f_{b,3402} = 0.0693 \pm 0.0068$ and $f_{b,5129} = 0.095 \pm 0.014$. 

\begin{figure}[h!]
	\centering
	\includegraphics[width=.75\textwidth]{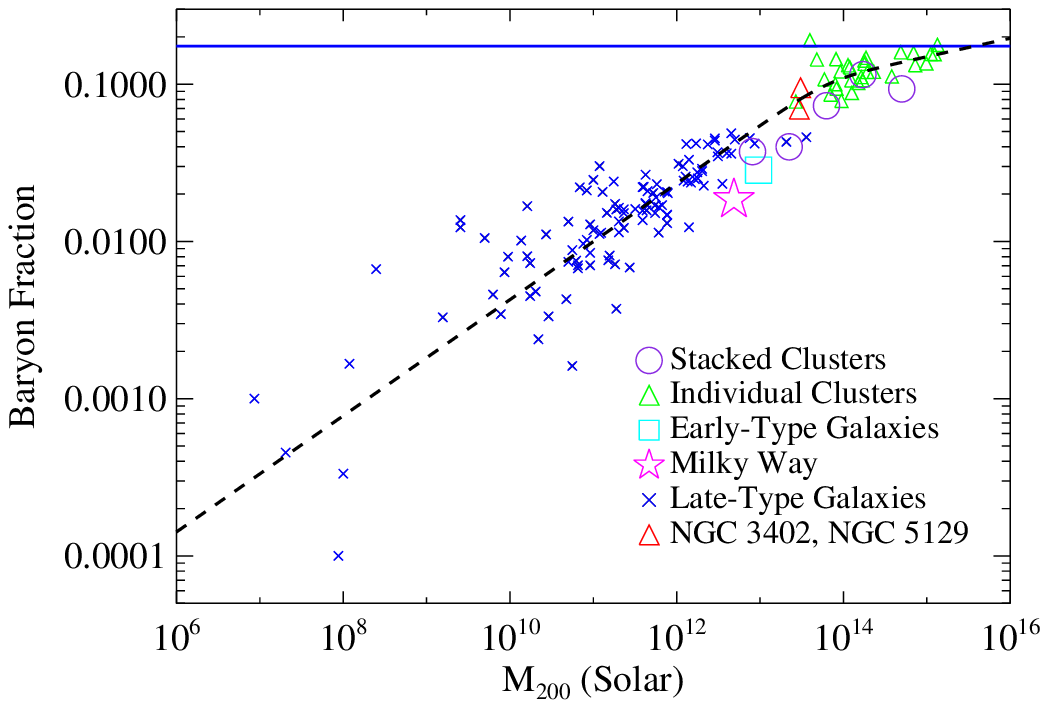}
	\caption{Baryon fraction as a function of $M_{200}$, or mass enclosed by $r_{200}$. Plotted are the measurements from \citet{saka03}, \citet{mcg05}, \citet{fly06}, \citet{vik06}, \citet{gav07}, \citet{walk07}, \citet{sta09}, \citet{sun09}, \citet{dai10}, \citet{and11}, and this work, converted from circular velocity to $M_{200}$. The blue solid line is the cosmological baryon fraction measured from the CMB, and the black dashed line is the best-fit broken power-law model for baryon losses.}
	\label{fig:bfrac}
\end{figure}

To compare our data with previous authors' work (Figure~\ref{fig:bfrac}), we first chose to convert Figure~10 in \citet{dai12} from the circular velocity ($V_{cir}$) at $r_{200}$ to the total gravitational mass enclosed within $r_{200}$ ($M_{200}$). This was done to provide a more intuitive representation of the physics. 
Here we used $M_{200}$ described in terms of the average mass density, $\rho_{ave}=200\,\rho_{crit}$, where $\rho_{crit}=3H^2(z)/8\pi G$ is the critical density of the universe. 
Since the objects in Figure~\ref{fig:bfrac} are relatively low redshift, we used $H(z)\approx H_0$. With this, we rewrote $M_{200}$ in terms of the circular velocity independent of $r$, \be M_{200}=\frac{V_{cir}^3}{10H_0 G}. \ee 
For our data, we generated $M_{200}$ by extrapolating Equation~\ref{eq:mtot2} out to $r_{200}$, which we computed from the 2$\beta$-model fits.
Next, we compared the $M_{200}$ estimates for stacked and individual clusters and our groups with the $M_{200}$--$T$ relation in Table~3 from \citet{dai07}, $M_{200}=Y_0(T/X_0)^k$, where $logY_0=13.58\pm0.05$, $X_0=1$ keV, and $k=1.65\pm0.12$. Many systems, including NGC 5129, had percent errors from the relation larger than 16\%. Thus, we utilized the $M_{200}$--$T$ relation to approximate the $M_{200}$ values for stacked and individual clusters, as well as our data. Then, we combined all data and fit with a broken power-law model of the same form as in \citet{dai10,dai12},
\be
f_b = \frac{0.109\,(M_{200}/6.41\times10^{13} M_{\odot})^a}{(1+(M_{200}/6.41\times10^{13} M_{\odot})^c)^{b/c}},
\ee
where $a = -0.369$, $b = 0.252$, and $c = 2$ (fixed at a smooth break). Above the break, the baryon fraction, $f_b$, scales as $f_b \propto M_{200}^{a-b=  -0.621}$ and $f_b \propto M_{200}^{a= -0.369}$ below the break. 
Figure~\ref{fig:bfrac} depicts the baryon fraction for all systems compiled in Figure~10 of \citet{dai12}, plus our data with the best-fit broken power-law model overlaid. Table~\ref{tab:fbmass} provides all mass components and baryon fractions for the two groups, as well as their emission-weighted radii in familiar overdensity forms. 
Also shown in Table~\ref{tab:fbmass} are the values determined for the baryon fractions out to $r_{500}$ and  $r_{200}$, along with another useful quantity, the gas fraction, $f_{gas}$. We derived the gas fraction for our groups out to $r_{emw}$, $r_{500}$, and  $r_{200}$. 

The extrapolated baryon fraction out to $r_{200}$ indicates a significant increase toward the cosmic value for NGC 3402. As for NGC 5129, it reached the cosmic fraction between $r_{500}$ and $r_{200}$. We further extrapolated the baryon fraction of NGC 3402 to $r_{100}$, the virial radius for the current cosmology. This resulted in $f_{b,100} = 0.184$, where the stellar, gas, and total masses are $M_{*,100} = 5.91 \times 10^{11}\ M_{\odot}$, $M_{gas, 100} = 6.81 \times 10^{12}\ M_{\odot}$, and $M_{tot, 100} = 4.03 \times 10^{13}\ M_{\odot}$, respectively. Thus, the $f_b$ overtook the cosmic fraction between $r_{200}$ and $r_{100}$ for NGC 3402. These findings strongly imply that much of the expected baryon content lies well outside $r_{500}$ but within the virial radii for these groups. Yet this is solely based on extrapolation and should not be overemphasized.

\begin{figure}[h!]
	\centering
	\includegraphics[width=.75\textwidth]{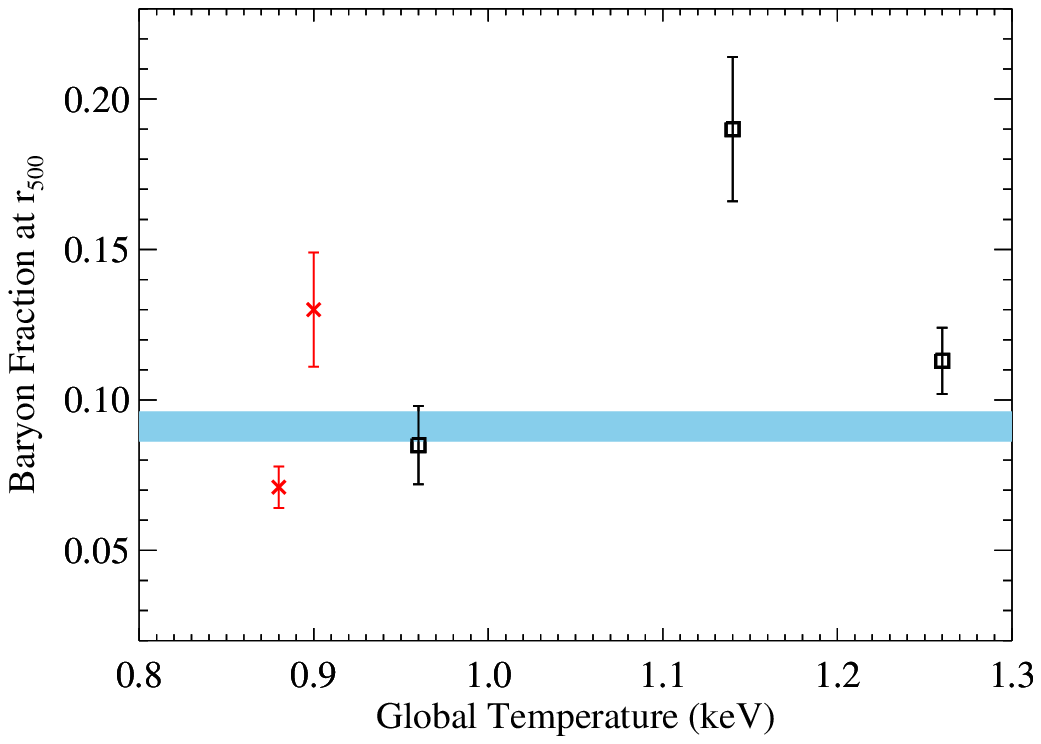}
	\caption{Baryon fraction out to $r_{500}$ vs. temperature, plotted for five galaxy groups with global gas temperatures less than 1.3 keV and whose baryon fractions were determined within $r\geq0.62\,r_{500}$. The blue filled region is the Bayesian averaged $f_b$ and 1$\sigma$ uncertainty, whereas the red crosses are the results from this work and black squares are the data from \citet{sun09}.}
	\label{fig:fbvst}
\end{figure}

To glean a further understanding of the baryon fractions of galaxy groups with low temperatures ($k_BT \lesssim 1.3$~keV) and measured at large radii, we combined our data with those of a previous work. There are three other groups, all from \citet{sun09}, whose adjusted global temperatures are measured out to a significant fraction of $r_{500}$. We combined their gas fractions extrapolated or measured out to $r_{500}$ with stellar estimates obtained using the redshift cutoff method used in Section~\ref{sec:mgas} to determine baryon fractions out to $r_{500}$. Listed in Table~\ref{tab:sungrps} are the $f_b$ at $r_{500}$, global $k_BT$ and measurement radii, where we symmetrized their uncertainties. Then, we plotted these groups with our extrapolated $f_{b,500}$ for NGC 3402 and NGC 5129 and computed the Bayesian average, $\overline{f_{b,500}} = 0.0912 \pm 0.0050$, which is shown in the blue filled region of Figure~\ref{fig:fbvst}. We have made the prior assumption of a Gaussian distribution for the mean $f_{b,500}$ being determined. The averaged $f_{b,500}$ falls significantly below the cosmological value for $\Omega_m = 0.26$ and $\Omega_\Lambda=0.74$, $f_{b,CMB} = 0.175 \pm 0.012$. For Planck 2018 cosmology, the cosmic baryon fraction is $f_{b,CMB} = 0.157$ \citep{pl2018}. 

We conclude that, on average, significant baryon deficits exist for poor groups within $r_{500}$ with temperatures between 0.8 and 1.3~keV.  
Other recent studies also found deficits of baryons in galaxy groups, although at higher temperatures of 2--3~keV \citep{lagan13, san13}.
These results reinforce our conclusion that the galaxy group regime is where baryon deficits become significant, insofar as the baryons were able to be detected. 
Through extrapolation of our mass estimates, we found that our poor groups most likely contain the cosmic proportion of baryons within the virial radius. However, this conclusion is hindered by the radial extent of our measurements, and future observations to even larger radii are needed to confirm this assessment.

This sample brings the X-ray community another step closer to understanding key differences between various galactic systems, which in turn should assist in constraining numerical simulations for both cosmology and the formation and evolution of these objects (e.g., \citealt{vik09}; \citealt{krav12}, and references therein; \citealt{hen18}). This includes identifying the mechanism for which the missing baryon problem occurs in different systems of galaxies. To achieve this, more outskirts observations close to the virial radius of similar systems are needed. 

\acknowledgements 
We thank E.\ O'Sullivan for providing the updated \xmm\ and \chan\ inner temperature profiles of NGC~3402 and for his helpful comments. Also, we appreciate Eric Miller for his assistance in the stray light issue and for answering several questions regarding the inner workings of \verb+Xspec+, \verb+xissim+, and \verb+xissimarfgen+. This research has made use of the NASA/IPAC Extragalactic Database (NED), which is operated by the Jet Propulsion Laboratory, California Institute of Technology, under contract with the National Aeronautics and Space Administration. In addition, this research has utilized data, software, and/or web tools obtained from NASA's High Energy Astrophysics Science Archive Research Center (HEASARC), a service of Goddard Space Flight Center and the Smithsonian Astrophysical Observatory. Also, we acknowledge support for this work from the NASA grants NNX11AG96G and NNX11AD09G and the NSF grant AST-1413056. This research has made use of data obtained from the \tele\ satellite, a collaborative mission between the space agencies of Japan (JAXA) and the USA (NASA).

\clearpage
\begin{deluxetable}{lccccccc}
	\tabletypesize{\scriptsize}
	\tablecolumns{8}
	\tablewidth{0pt}
	\tablecaption{Observation Parameters}
	\tablehead{
		\colhead{Observation}  & \colhead{ObsID} & \colhead{Date} & \colhead{R.A. (J2000)} & \colhead{Decl. (J2000)} & \multicolumn{3}{c}{Cleaned/Final Exposure Time (ks)\tablenotemark{a}}
		\\
		\colhead{}  & \colhead{} & \colhead{} & \colhead{(deg)} & \colhead{(deg)} & \colhead{XIS0} & \colhead{XIS1} & \colhead{XIS3} 
	}
	\startdata
	NGC 3402 & 805070010 & 2010 Dec 27 & 162.4923 & $-$13.1954 & 27.9/24.1 & 27.9/24.1 & 27.9/24.1 \\
	NGC 3402 back & 805071010 & 2010 Dec 19 & 161.6656 & $-$13.5535 & 12.5/12.5 & 12.5/12.5 & 12.5/12.5 \\
	NGC 5129 1st & 805072010 & 2010 Dec 18 & 201.3141 & 14.0346 & 30.3/25.6 & 30.3/25.6 & 30.3/25.6 \\
	NGC 5129 2nd & 805073010 & 2010 Dec 18 & 201.1253 & 13.7341 & 30.8/25.4 & 30.8/25.3 & 30.8/25.4 \\
	NGC 5129 back & 805074010 & 2010 Dec 17 & 201.7433 & 13.5725 & 12.3/12.3 & 12.3/12.3 & 12.3/12.3 \\
	\enddata
	\label{tab:obspar}
	\tablenotetext{a}{Cleaned exposure times resulting from routine screening, whereas final exposure times are after all screening, including the COR2 $>$ 6 GV condition.}
\end{deluxetable}

\begin{deluxetable}{cccccc}
	\tabletypesize{\scriptsize}
	\tablecolumns{6}
	\tablewidth{0pt}
	\tablecaption{Xspec Background Parameters and Normalizations for Spectral Analysis}
	\tablehead{
		\colhead{Emission Source}  & \colhead{Model Type} & \colhead{Parameter} & \colhead{Fixed/Free} & \colhead{NGC 3402} & \colhead{NGC 5129} \\
		\colhead{or Absorption} & \colhead{} & \colhead{} & \colhead{} & \colhead{} & \colhead{}
	}
	\startdata
	Galactic absorption & wabs & $N_H (10^{22}$ cm$^{-2})$ & Fixed  & 0.0477 & 0.0178 \\
	AGN & power-law & $\Gamma$ & Fixed & 1.41 & 1.41 \\
	& & Normalization\tablenotemark{a} & Free & $(7.63\pm0.23)\e{-4}$ & $(1.05 \pm 0.03) \e{-3}$\\
	Galaxy & apec & $k_BT$ (keV) & Free & $0.177 \pm 0.007$ & $0.173 \pm 0.009$ \\
	& & Abundance ($Z_{\odot}$) & Fixed & 1 &  1 \\
	& & Redshift & Fixed & 0 & 0 \\
	& & Normalization\tablenotemark{b} & Free & $(1.65 \pm 0.19) \e{-3}$ & $(3.49 \pm 0.29) \e{-3}$ \\
	Galaxy (NPS)\tablenotemark{c} & apec & $k_BT$ & Fixed &  \nodata & 0.4 \\
	& & Abundance & Fixed &  \nodata  & 1 \\
	& &  Redshift & Fixed &  \nodata & 0 \\
	& & Normalization\tablenotemark{b} & Free &  \nodata & $(6.28 \pm 1.20) \e{-4}$\\
	\enddata
	\label{tab:bkgdpar}
	\tablecomments{All normalizations assume an emission area of $400\pi$ in Xspec.}
	\tablenotetext{a}{Power-law normalization in photons cm$^{-2}$ s$^{-1}$ keV$^{-1}$ at 1 keV.}
	\tablenotetext{b}{\emph{Apec} normalization given in cm$^{-5}$.}
	\tablenotetext{c}{Excess emission due to the North Polar Spur (NPS).}
\end{deluxetable}

\begin{deluxetable}{ccccccc}
	\tabletypesize{\scriptsize}
	\tablecolumns{7}
	\tablewidth{0pt}
	\tablecaption{Model Emission Lines and Their Candidates}
	\tablehead{
		\colhead{Observation}  & \multicolumn{3}{c}{FI CCDs} & \multicolumn{3}{c}{BI CCD} \\
		\colhead{} & \colhead{Line Energy} & \colhead{Candidate} & \colhead{Emission Type} & \colhead{Line Energy} & \colhead{Candidate} & \colhead{Emission Type} \\
		\colhead{} & \colhead{keV} & \colhead{} & \colhead{} & \colhead{keV} & \colhead{} & \colhead{}
	}
	\startdata
	NGC 3402 & 1.08 & Ne X Ly$\alpha$ & SWCX & 1.48 & Al K$\alpha$ & NXB \\
	\tableline 
	\noalign{\vskip 0.15cm}
	NGC 3402 Back & 0.685 & O VII & SWCX & 0.58 & O VII K$\alpha$ & SWCX \\
	& 0.815 & O VIII Ly$\gamma$ & SWCX & 1.285 & Ne X & SWCX \\
	& 1.825 & Si XIII & SWCX & 2.195 & Au M$\alpha$ & NXB \\
	\tableline 
	\noalign{\vskip 0.15cm}
	NGC 5129 & 0.65 & O VIII Ly$\alpha$ & SWCX &  0.55 & O VII K$\alpha$ & SWCX \\ 
	& 0.915 & Ne IX K$\alpha$ & SWCX & 0.85 & O VIII Ly$\epsilon$ & SWCX \\
	\tableline 
	\noalign{\vskip 0.15cm}
	NGC 5129 Back & 0.63 & N VII & SWCX & 0.665 & O VII K$\beta$ & SWCX \\
	& 0.915 & Ne IX K$\alpha$ & SWCX & 0.805 & O VIII Ly$\gamma$ & SWCX \\
	& 1.37 & Mg XI K$\alpha$ & SWCX & 0.895 & Ne IX K$\alpha$ & SWCX \\
	& 2.155 & Au M$\alpha$ & NXB &  &  & \\
	\enddata
	\label{tab:lines}
	\tablecomments{All line energies are the centers of residual lines as observed by the \suz~XISs.}
\end{deluxetable}

\begin{deluxetable}{cccccc}
	\tabletypesize{\scriptsize}
	\tablecolumns{6}
	\tablewidth{0pt}
	\tablecaption{Xspec Group Parameters and Normalizations for Spectral Analysis}
	\tablehead{
		\colhead{Emission Source}  & \colhead{Model Type} & \colhead{Parameter} & \colhead{Fixed/Free} & \colhead{NGC 3402} & \colhead{NGC 5129} \\
		\colhead{or Absorption} & \colhead{} & \colhead{} & \colhead{} & \colhead{} & \colhead{}
	}
	\startdata
	Galactic absorption & wabs & $N_H (10^{22}$ cm$^{-2})$ & Fixed  & 0.046 & 0.0176 \\
	Group hot halo & apec & $k_BT$ (keV) & Free & 0.862$\substack{+0.093\tablenotemark{a}\\-0.112}$ $\pm$ 0.054\tablenotemark{b}&0.962 $\substack{+0.215\tablenotemark{a}\\-0.147}$$\pm$ 0.066\tablenotemark{b} \\
	& & Abundance ($Z_{\odot}$) & Fixed & 0.2 & 0.2  \\
	& & Redshift & Fixed &  0.0153 & 0.0230 \\ 
	& & Normalization\tablenotemark{c} & Free & 5.24$\substack{+1.29\tablenotemark{a}\\-1.15}$$\pm$ 0.76\tablenotemark{b} & 5.09 $\pm 1.91$\tablenotemark{a} $\pm$  1.11\tablenotemark{b} \\
	& &$\chi_{min}^2/dof$ &  & 249/317 & 471/506 \\
	& &$\chi_{\nu}^2$ range\tablenotemark{d} & & 0.784--0.894 & 0.917--1.06 \\
	\enddata
	\label{tab:grouppar}
	\tablecomments{All normalizations assume an emission area of $400\pi$ in Xspec.}
	\tablenotetext{a}{Statistical uncertainties}
	\tablenotetext{b}{Systematic uncertainties based on the eight different background models discussed in Section~\ref{ssec:grmod}.}
	\tablenotetext{c}{\emph{Apec} normalization given in $10^{-4}$ cm$^{-5}$.}
	\tablenotetext{d}{Range in reduced $\chi^2$ for the eight different background and group simultaneous models.}
	
\end{deluxetable}

\begin{deluxetable}{lccc}
	\tabletypesize{\scriptsize}
	\tablecolumns{4}
	\tablewidth{0pt}
	\tablecaption{Mean Surface Brightnesses and Electron Number Densities at $R_{emw}$}
	\tablehead{
		\colhead{Spectral Analysis} & \colhead{FI/BI} & \multicolumn{2}{c}{Observations} \\
		&	& \colhead{NGC 3402} & \colhead{NGC 5129}   
	}
	\startdata
	\colhead{$S_{0.6-1.3}$ (10$^{-8}$ counts~s$^{-1}$~kpc$^{-2}$)\tablenotemark{a}\tablenotemark{b}} & FI/BI & 3.04 \pum~0.71 / 5.4 \pum~1.3 & 1.36 \pum~0.51 / 2.41 \pum~0.90\\
	\colhead{$S_{0.7-2.0}$ (10$^{-8}$ counts~s$^{-1}$~kpc$^{-2}$)\tablenotemark{a}\tablenotemark{c}} & FI/BI & 18.0 \pum~4.2 / 17.6 \pum~4.1 & 6.3 \pum~2.4 / 6.2 \pum~2.3 \\
	\colhead{$\Sigma_{0.1-2.0}$ (10$^{-10}$ photons~s$^{-1}$~cm$^{-2}$~arcsec$^{-2}$)\tablenotemark{d}} & $\cdots$ & 2.16 \pum~0.53  & 1.91 \pum~0.77 \\
	\colhead{$n_e$ (10$^{-5}$ cm$^{-3}$)} & $\cdots$ & 6.55 \pum~0.79 & 14.6 \pum~2.8 \\
	\enddata
	\label{tab:sbnd}
	\tablenotetext{a}{Note that the effective areas of the XISs have not been divided, since they are energy dependent.}
	\tablenotetext{b}{This SB is in the 0.6--1.3 keV band for the \suz~observations.}
	\tablenotetext{c}{The CR used to create this SB has been converted to \chan~in the 0.7--2.0 keV band.}
	\tablenotetext{d}{To match \citet{eck11}, this SB is in the 0.1--2.0 keV band.}
	
\end{deluxetable}

\begin{deluxetable}{ccc}
	\tabletypesize{\scriptsize}
	\tablecolumns{6}
	\tablewidth{0pt}
	\tablecaption{2$\beta$-Model Fit Parameters}
	\tablehead{
		\colhead{Model Parameters}  & \multicolumn{2}{c}{Value (inner plus our data)}
		\\
		\colhead{} & \colhead{NGC 3402} & \colhead{NGC 5129}
	}
	\startdata
	$r_{c1}$(kpc) & 2.61 $\pm$ 0.11 & 14.08 \\
	$\beta_{1}$ & 0.4925 $\pm$ 0.0037 & 1.432 $\pm$ 0.074 \\
	$\Sigma_{01}$(photons~s$^{-1}$~cm$^{-2}$~arcsec$^{-2}$) & (1.610 $\pm$ 0.078) $\times$ 10$^{-6}$ & (9.22 $\pm$ 0.46) $\times$ 10$^{-8}$ \\
	$r_{c2}$(kpc) & 138.4 & 38.4 $\pm$ 6.8 \\
	$\beta_{2}$ & 4.55 $\pm$ 0.29 & 0.507 $\pm$ 0.038 \\
	$\Sigma_{02}$ (photons~s$^{-1}$~cm$^{-2}$~arcsec$^{-2}$) & (1.30 $\pm$ 0.12) $\times$ 10$^{-8}$ & (8.60 $\pm$ 0.97) $\times$ 10$^{-9}$ \\
	$\chi_{min}^2/$dof & 56.1/43 & 26.6/16 \\
	\enddata
	\tablecomments{Best-fit parameters for the 2$\beta$-model considering inner \chan\ data and our \suz\ contributions. Note that the dof could be as high as 47 and 20, respectively.}
	\label{tab:2betafit}
\end{deluxetable}

\begin{deluxetable}{ccc}
	\tabletypesize{\scriptsize}
	\tablecolumns{3}
	\tablewidth{0pt}
	\tablecaption{Derived Group Properties}
	\tablehead{
		\colhead{Property} & \colhead{NGC 3402} & \colhead{NGC 5129}
	}
	
	\startdata
	$M_{*, emw}$ ($10^{11} M_{\odot}$) & $2.87\pm0.43$ & $7.11\pm0.96$\\
	$M_{*, 500}$ ($10^{11} M_{\odot}$) & 2.87 & 7.99\\
	$M_{*, 200}$ ($10^{11} M_{\odot}$) & 4.12 & 10.6\\
	$M_{gas, emw}$($10^{11} M_{\odot}$) & $9.3\pm1.1$ & $6.1\pm1.2$ \\
	$M_{gas, 500}$($10^{11} M_{\odot}$) & 9.9 & 19 \\
	$M_{gas, 200}$($10^{11} M_{\odot}$) & 30 & 46 \\
	$M_{tot, emw}$($10^{13} M_{\odot}$) & $1.750\pm0.013$ & $1.39\pm 0.12$ \\
	$M_{tot, 500}$($10^{13} M_{\odot}$) & 1.80 & 2.05 \\
	$M_{200, M-T}$\tablenotemark{a}($10^{13} M_{\odot}$) & 2.95 & 3.06 \\
	$M_{tot, 200}$($10^{13} M_{\odot}$) & 2.85 & 2.63 \\
	\tableline
	\noalign{\vskip 0.15cm}
	$f_{g, emw}$ & $0.0530\pm0.0063$ & $0.0438\pm0.0091$ \\
	$f_{g, 500}$ & 0.0551 & 0.0911 \\
	$f_{g, 200}$ & 0.104 & 0.175 \\
	$f_{b, emw}$ & $0.0693\pm0.0068$ & $0.095\pm 0.014$ \\
	$f_{b, 500}$ & 0.071 & 0.13 \\
	$f_{b, 200}$ & 0.118 & 0.216 \\
	\tableline
	\noalign{\vskip 0.15cm}
	$r_{emw}$(kpc)\tablenotemark{b} & 375 & 249\\
	$\Delta$\tablenotemark{c} & 530 & 1430\\
	$r_{500}$(kpc)  & 386 & 402 \\
	$r_{200}$(kpc)  & 610 & 593\\
	$K_{emw}$ (keV~cm$^2$) & $530\pm76$ & $348\pm79$\\
	\tableline
	\noalign{\vskip 0.15cm}
	$L_{X,bol, emw}$($10^{42}$~\lumin) & $7.00\pm0.15$ & $3.157\pm0.074$ \\
	$L_{X,bol, 500}$($10^{42}$~\lumin) & 7.02 & 3.23 \\
	$L_{X,bol, 200}$($10^{42}$~\lumin) & 7.21 & 3.35 \\
	$L_{ROSAT}$($10^{42}$~\lumin) & $6.77\pm0.15$ & $3.042\pm0.071$ \\
	$F_{X}$($10^{-12}$~\flux) & $9.09\pm0.20$ & $1.790\pm0.042$ \\
	\enddata
	\tablecomments{All quantities derived are based on $h = 0.73$ and are related to the Hubble constant by $M_*\propto h^{-2}$, $M_{g}\propto h^{-5/2}$, $M_{tot}\propto h^{-1}$, $L_X\propto h^{-2}$, $\Sigma\propto h^{-1/3}$, and $r\propto h^{-1}$.}
	\label{tab:fbmass}
	\tablenotetext{a}{This value for $M_{200}$ was derived from the Poisson fit to the $M_{200}$--$T$ relation in \citet{dai07}.}
	\tablenotetext{b}{The $r_{emw}$ here is the emission-weighted radius.}
	\tablenotetext{c}{$\Delta$ is the constant term that when multiplied by $\rho_{crit}$ gives the average mass density of the group.}
\end{deluxetable}

\begin{deluxetable}{cccc}
	\tabletypesize{\scriptsize}
	\tablecolumns{4}
	\tablewidth{0pt}
	\tablecaption{\citet{sun09} Groups and Properties}
	\tablehead{
		\colhead{Galaxy Group}  & \colhead{$r_{obs}/r_{500}$} & \colhead{$f_b$} & \colhead{$k_BT$ (keV)}}
	\startdata
	NGC 1550 & 0.76 & $0.113 \pm 0.011$ & $1.26 \pm 0.02$ \\
	NGC 5098 & 1.06 & $0.190 \pm 0.024$ & $1.14 \pm 0.05$ \\
	UGC 5088 & 0.87 & $0.085 \pm 0.013$ & $0.96 \pm 0.04$ \\
	\enddata
	\tablecomments{Properties of the groups measured out to or near $r_{500}$ in \citet{sun09}, adjusted for the change in AtomDB.}
	\label{tab:sungrps}
\end{deluxetable}

\end{document}